\documentclass[10pt,journal,compsoc]{IEEEtran}
\ifCLASSOPTIONcompsoc
  \usepackage[nocompress]{cite}
\else
  \usepackage{cite}
\fi

\usepackage{amsmath,amssymb,amsfonts,chemarrow,balance}
\usepackage{graphicx}
\usepackage{pifont}
\newcommand{\cmark}{\ding{51}}%
\newcommand{\xmark}{\ding{55}}%

\usepackage[inline]{enumitem}
\usepackage{mathenv}
\usepackage[english]{babel}
\usepackage{breqn}
\newcommand{\BfPara}[1]{\vspace{1mm}{\noindent\bf#1.}\xspace\xspace}
\usepackage{url}
\usepackage{subfigure}
\usepackage{multirow}

\usepackage[linesnumbered,ruled]{algorithm2e}
\usepackage{algpseudocode}
\usepackage{booktabs}
\usepackage{graphics}
\usepackage{epsfig}
\usepackage{listings}
\usepackage{rotating}
\usepackage{amsthm}
\usepackage{hyperref}
\usepackage{xcolor}
\usepackage{upquote}
\usepackage{xspace}
\usepackage{setspace}

\newcommand{\note}[1]{}    
\newcommand{\slx}{{\em Selenium}\xspace}

\newcommand{\pl}{{\em Plato}\xspace}

\newcommand{\etc}{{etc.}\xspace}

\newcommand{\ie}{{\em i.e.,}\xspace}

\newcommand{\Cc}{{Cryptocurrency}\xspace}

\newcommand{\cj}{cryptojacking\xspace}

\newcommand\JSONstringvaluestyle{\color{red}}

\newcommand{\tsref}[1]{\textsection\ref{#1}\xspace}

\newcommand{\etal}{{\em et al.}\xspace}

\colorlet{punct}{red!60!black}
\definecolor{background}{HTML}{ffffff }
\definecolor{delim}{RGB}{20,105,176}
\colorlet{numb}{magenta!60!black}
\definecolor{light-gray}{gray}{0.95}
\definecolor{darkgray}{rgb}{0.4, 0.4, 0.4}
\definecolor{editorGray}{rgb}{0.95, 0.95, 0.95}
\definecolor{editorOcher}{rgb}{1, 0.5, 0} 
\definecolor{editorGreen}{rgb}{0, 0.5, 0} 
\definecolor{orange}{rgb}{1,0.45,0.13}      
\definecolor{olive}{rgb}{0.17,0.59,0.20}
\definecolor{brown}{rgb}{0.69,0.31,0.31}
\definecolor{purple}{rgb}{0.38,0.18,0.81}
\definecolor{lightblue}{rgb}{0.1,0.57,0.7}
\definecolor{lightred}{rgb}{1,0.4,0.5}

\lstdefinelanguage{JavaScript}{
  morekeywords={typeof, new, true, false, catch, function, return, null, catch, switch, var, if, in, while, do, else, case, break},
  morecomment=[s]{/*}{*/},
  morecomment=[l]//,
  morestring=[b]",
  morestring=[b]'
}
\SetKwFor{For}{for (}{) $\lbrace$}{$\rbrace$}
\lstdefinelanguage{HTML5}{
  language=html,
  sensitive=true,   
  alsoletter={<>=-:},    
  morecomment=[s]{<!-}{-->},
  tag=[s],
  otherkeywords={
  >,
    <!DOCTYPE,
  </html, <html, <head, <title, </title, <style, </style, <link, </head, <meta, />,
    </body, <body,
    </div, <div, </div>, 
    </p, <p, </p>,
    </script, <script,
  <canvas, /canvas>, <svg, <rect, <animateTransform, </rect>, </svg>, <video, <source, <iframe, </iframe>, </video>, <image, </image>, <header, </header, <article, </article
  },
  ndkeywords={
  =,
  charset=, src=, throttle:, id=, width=, height=, style=, type=, rel=, href=,
  fill=, attributeName=, begin=, dur=, from=, to=, poster=, controls=, x=, y=, repeatCount=, xlink:href=,
  margin:, padding:, background-image:, border:, top:, left:, position:, width:, height:, margin-top:, margin-bottom:, font-size:, line-height:,
  transform:, -moz-transform:, -webkit-transform:,
  animation:, -webkit-animation:,
  transition:,  transition-duration:, transition-property:, transition-timing-function:,
  }
}

\usepackage{pgfplots}
\usepackage{pgfplotstable}

\lstdefinestyle{htmlcssjs} {%
    backgroundcolor=\color{light-gray},
    basicstyle=\footnotesize\ttfamily,
    showstringspaces=false,
    frame = lines, 
    breaklines=true,
    showstringspaces =false,
    keywords = {false,true},
  backgroundcolor=\color{light-gray},
  identifierstyle=\color{black},
  keywordstyle=\color{blue},
  ndkeywordstyle=\color{black},
  stringstyle=\color{red}\ttfamily,
  commentstyle=\color{brown}\ttfamily,
  language=HTML5,
  alsolanguage=JavaScript,
  alsodigit={.:;},  
  tabsize=2,
  showtabs=false,
  showspaces=false,
  showstringspaces=false,
  extendedchars=true,
  breaklines=true,
  literate=%
  {Ã}{{\"O}}1
  {Ã}{{\"A}}1
  {Ã}{{\"U}}1
  {Ã}{{\ss}}1
  {ÃŒ}{{\"u}}1
  {Ã€}{{\"a}}1
  {Ã¶}{{\"o}}1
}

\lstdefinestyle{json}
{ backgroundcolor=\color{light-gray},
    basicstyle=\footnotesize\ttfamily,
    showstringspaces=false,
    breaklines=true,
    frame=lines,
  showstringspaces    =false,
  keywords            = {false,true},
  commentstyle= \itshape\color{codegreen},
  alsoletter          =0123456789.,
  morestring          = [s]{"}{"},
  stringstyle         = \ifcolonfoundonthisline\JSONstringvaluestyle\fi,
  MoreSelectCharTable =%
    \lst@DefSaveDef{`:}\colon@json{\processColon@json},
  keywordstyle        = \ttfamily\bfseries,
}

\SetKwRepeat{Do}{do}{while}%

\def\equationautorefname~#1\null{(#1)\null}
\newif\ifcolonfoundonthisline
\makeatletter

\newcommand\processColon@json{%
  \colon@json%
  \ifnum\lst@mode=\lst@Pmode%
    \global\colonfoundonthislinetrue%
  \fi
}

\usepackage{pgfplots} 
\usetikzlibrary{patterns}

\begin{document}

\title{Analyzing In-browser Cryptojacking}

\author{Muhammad Saad and David Mohaisen~\IEEEmembership{Senior Member,~IEEE.}
\thanks{M. Saad, and D. Mohaisen are with PayPal and the University of Central Florida. M.Saad is the corresponding author. E-mail: muhsaad@paypal.com. An earlier version of this work has appeared in APWG Symposium on Electronic Crime Research \cite{SaadKM19}}}


\IEEEtitleabstractindextext{
\begin{abstract}
Cryptojacking is the permissionless use of a target device to covertly mine cryptocurrencies. With cryptojacking, attackers use malicious JavaScript codes to force web browsers into solving proof-of-work puzzles, thus making money by exploiting the resources of the website visitors. To understand and counter such attacks, we systematically analyze the static, dynamic, and economic aspects of in-browser cryptojacking. For static analysis, we perform content, currency, and code-based categorization of cryptojacking samples to 1) measure their distribution across websites, 2) highlight their platform affinities, and 3) study their code complexities. We apply machine learning techniques to distinguish cryptojacking scripts from benign and malicious JavaScript samples with 100\% accuracy. For dynamic analysis, we analyze the effect of cryptojacking on critical system resources, such as CPU and battery usage. We also perform web browser fingerprinting to analyze the information exchange between the victim node and the dropzone cryptojacking server. We also build an analytical model to empirically evaluate the feasibility of cryptojacking as an alternative to online advertisement. Our results show a sizeable negative profit and loss gap, indicating that the model is economically infeasible. Finally, leveraging insights from our analyses, we build countermeasures for in-browser cryptojacking that improve the existing remedies.  
\end{abstract}
}
\maketitle

\section{Introduction}\label{sec:introduction}
Blockchain-based cryptocurrencies have emerged as an innovation in distributed systems, enabling a transparent and distributed storage of transactions. Various proof mechanisms, such as the Proof-of-Work (PoW), prevent abuse and improve cryptocurrency trustworthiness. In Bitcoin, for example, individual miners mine new coins through extensive hash operations, which are then verified by distributed nodes in a peer-to-peer (P2P) network~\cite{SaadARM21,SaadCM21b,SaadCM21}. However, PoW led to abuse: an adversary may employ various techniques to abuse public resources for mining purposes and to perform extensive hash calculations at no or low cost. 

Cryptojacking is the use of resources of a target host to compute hashes and make a profit out of mining without the consent of the target's owner. Conventional cryptojacking involved the installation of a software binary on a target host that secretly solved PoW and communicated the results to a remote server~\cite{Scott_18}. Such conventional cryptojacking required user permission to download the software and a persistent Internet connection to communicate the PoW result to the adversary or a {\em dropzone} server controlled by him. However, conventional cryptojacking proved infeasible for several reasons. First, not all devices have a persistent Internet connection when needed to send PoW results; PoWs are time-sensitive. If not sent immediately after being solved, PoWs become easily outdated. Secondly, antivirus companies can easily identify binaries used for cryptojacking and detect them~\cite{Zuckerman_18}. Finally, this attack requires an infection vector, whereby users would enable the attack by mistakenly installing the cryptojacking binaries. 

A recent form of in-browser cryptojacking that does not suffer from those issues has emerged. In-browser cryptojacking does not require installing binaries or authorization from users to operate. In-browser cryptojacking instances use {\em JavaScript} code to compute PoW in the web browser and transmit the PoW to a remote server~\cite{Slm_18}. As such, and since they are shielded in the browser's process, they are undetected by the antivirus scanners. Moreover, mining during browsing  ensures uninterrupted transmission of PoW over a persistent Internet connection. 

Initially intended for good use as an alternative revenue source to online advertisement~\cite{kerbs}, in-browser cryptojacking was made easy by online services such as {\em Coinhive}~\cite{coinhive}, which provided {\em JavaScript} templates for cryptojacking. {\em Coinhive} provided scripts to mine Monero. This cryptocurrency is hard to trace and rewards miners based on the aggregated hashes they contributed. It has been a major concern to users as evidenced by the trend in search engines and by security operation companies pointing out to their significant growth. For instance, in August 2020, Symantec reported a 163\% increase in cryptojacking attacks in the second quarter of 2020~\cite{teamsymantec20}. Their report on the threat landscape~\cite{teamsymantec20} identifies a correlation between the cryptocurrency price and the cryptojacking events, acknowledging that cryptojacking is still a threat in the web ecosystem. 

In-browser cryptojacking serves as an attack avenue for hackers who inject malicious {\em JavaScript} scripts into popular websites without the knowledge of website owners and mine cryptocurrency for themselves. According to Symantec's Security Threat Report (ISTR), cryptojacking attacks on websites rose by 8500\% during 2017~\cite{Mathur_18,Singh_18}. In February 2018, a major cryptojacking attack hit more than 4000 websites worldwide, including the websites of the US Federal Judiciary and the UK National Health Service (NHS)~\cite{condliffe_18}. Also, in February 2018, Tesla became the victim of a cryptojacking attack in which attackers hijacked Tesla cloud and deployed their cryptojacking code~\cite{Rayome_18}. After such unusual incidents, UK's National Cyber Security Centre (NCSC) indicated cryptojacking as a ``significant threat'' in its latest cyber security report~\cite{de_18,ncsc_18}. 

The use of cryptojacking as a replacement for advertisement also has witnessed a great debate. For example, some popular websites, such as ``The Pirate Bay'', started using cryptojacking as a revenue substitute to online advertisement~\cite{Shaikh_17,Ernesto_17,Jones_2017}. ``The Pirate Bay'' website later disclosed to its users that it would use the CPU cycles of the visitors in exchange for ad-free web browsing, garnering users' approval. As some other websites started using cryptojacking as a revenue generation mechanism, a debate was sparked on the ethics of using cryptojacking~\cite{Zuckerman_2018} and the absence of user consent. Furthermore, it was observed that the continuous CPU-intensive mining, especially on battery-powered devices, resulted in the quick drainage of those devices, adding a new variable to the debate of whether cryptojacking is a good alternative to online advertising. 

Motivated by these events, we conduct an in-depth study on in-browser cryptojacking and its effects on website visitors and their devices. We start by analyzing and characterizing more than 5,700 websites with cryptojacking scripts. We then explore static and dynamic analysis tools to understand the behavioral traits of in-browser cryptojacking scripts toward their detection. Using various features extracted through this analysis, we build a classifier for detecting cryptojacking scripts among benign scripts, as well as other malicious types of {\em JavaScript} codes. We also measure the impact of in-browser cryptojacking on user devices regarding CPU usage and battery drainage. Finally, in examining the feasibility of cryptojacking as an alternative to online advertisement, we conduct an in-depth end-to-end analysis that considers the implications of such an alternative on both users and websites.

\BfPara{Contributions and Roadmap}
We make the following major contributions:
\begin{enumerate*}
    \item Using more than 5,700 websites with cryptojacking scripts, we conduct an in-depth analysis and characterization of cryptojacking (\textsection\ref{sec:prelim}). 
    \item We conduct static analysis of the cryptojacking scripts to highlight distributions of cryptocurrency used in cryptojacking and code (script) complexity analysis (\textsection\ref{sec:static}). We further built machine learning models to automatically classify cryptojacking (\textsection\ref{sec:clustering}). Our machine learning models achieved an accuracy of $\approx100\%$.
    \item We performed dynamic analysis to highlight the unique characteristics of process usage, battery usage, and dynamically generated data analysis through WebSocket inspection of cryptojacking scripts (\textsection\ref{sec:dynamic}). 
    \item We examine the economic arguments  for cryptojacking as an alternative to online advertisement and build a model to estimate the cost of cryptojacking to the users as well as the gain to websites conducting cryptojacking (\textsection\ref{sec:economic}). We show the economic model is impractical for benign use, and unprofitable for malicious use. 
    \item We explore the limitations of existing countermeasures and suggest new defenses (\textsection\ref{sec:counter}). 
    \end{enumerate*}
    
Atop of the contributions in~\cite{SaadKM19}, this paper extends our analysis by: 
    \begin{enumerate*}
    \item a more systematic and in-depth background about various aspects of cryptojacking including its prevalence, popularity, and association with blockchain-based cryptocurrencies, 
    
    \item adapting a supervised learning approach in which we used logistic regression, linear discriminant analysis, k-nearest neighbors, support vector machine, and random forest to improve the detection accuracy of cryptojacking codes (at the website level), and 
    \item analyzing the memory footprints of in-browser cryptojacking.
    \end{enumerate*}
 
To show that cryptojacking is still relevant, we  identify 620 out of the 5703 websites that are still active and revise our results in~\tsref{sec:prelim} and \tsref{sec:static}. Our new dataset and recent reports~\cite{teamsymantec20} show that cryptojacking is still a problem. 

The rest of the paper includes a background in \textsection\ref{sec:background}, the related work in~\textsection\ref{sec:rw}, and conclusion in~\textsection\ref{sec:conclusion}, respectively. 

\section{Background}\label{sec:background}

\subsection{Blockchain-based Cryptocurrencies}\label{sec:cryptos}
In 2009, the first blockchain-based digital currency ``Bitcoin'' was introduced by Satoshi Nakamoto~\cite{nakamoto2008bitcoin} that involved exchange of transactions without the use of a central authority~\cite{SaadSNKSNM20}. In Bitcoin, the role of the trusted central authority was replaced by a transparent and tamper-proof public blockchain that acted as a public ledger to maintain the records of transactions.  The consensus in the decentralized peer-to-peer Bitcoin network was augmented by cryptographically secure algorithm known as the proof-of-work (PoW). Bitcoin remained the only cryptocurrency for two years after which several more digital currencies joined the market. As of today, there are more than 5000 cryptocurrencies have been introduced in the market~\cite{atozforex} with more than 5.8 million active users~\cite{hileman2017global}. Bitcoin is leading the cryptocurrency market with a 58\% market share, or $\approx$\$4.9 Billion USD trade volume and more than 12,000 transactions per hour~\cite{bitcoinnews_2017}. 
Towards the end of 2016, the price of 1 bitcoin was a littler under \$1000 USD and during 2017 it rose to a market price of \$19,000 USD~\cite{blockexplorer}. Some other notable \Cc that make use of public blockchain are Ethereum, Litecoin, Ripple, Monero, and Dash. 

\subsection{Mining in Cryptocurrencies}\label{sec:mining}
The key operations in every cryptocurrency involve the exchange of transactions among peers, the mining of transactions in blocks, and the publishing of blocks containing those transactions. Computing a valid block results in the generation of new coins in the system.

However, computing a valid block is a non-trivial process in which miners must solve mathematical challenges and provide a PoW for their solutions. In Bitcoin, PoW involves finding a \textit{nonce} that, when hashed with the data in the block, produces a hash value less than the target threshold the system sets. The target is a function of network difficulty and is denoted by a 256-bit unsigned integer encoded in a 32-bit ``compact'' form and stored in the block header. In solving the challenge, miners spend effort and, in return, get rewarded with 12.5 bitcoins for each valid PoW. As more miners join, the hash power of the network and the probability of computing a block increase. The network's difficulty is adjusted every two weeks to keep the average block computation time within the fixed range (2016 blocks).

We show how the block computation time, $T(B)$, is affected by the hashing rate, $H_r$, the {\em target}, $Target$, the probability of finding a block, $P_r(B)$, and the average number of hashes required to solve the target, $H$. To keep  $T(B)$ in a fixed range (10 minutes), as the $H_r$ increases, the target value is adjusted to keep $P_r(B)$ constant. As such, we calculate $P_r(B)  = {Target}/{2^{256}}$,  $H = {1}/{P_r(B)}$, and $T(B) = {H}/{H_r} = {1}/({P_r(B)\times H_r})$.

\subsection{Cryptojacking}\label{sec:cj} 

Generally, attackers utilize two main strategies for unauthorized use of a victim's machine to mine digital currencies through cryptojacking:  installing a binary on the machine or using an in-browser script. The first one loads the mining code on the victim's machine as a stand-alone binary (or an infection of a binary). As such, it requires information about the target machine, including its operating system and hardware constructs. For example, a malicious cryptojacking binary developed for Windows cannot be executed on Linux. However, the second strategy is platform agnostic. The cryptojacking~{\em JavaScript} is executed upon loading the website in the victim's browser. In both cases, the mining code works in the background. Below, we briefly discuss the two cryptojacking strategies. However, the main focus of this paper is the in-browser cryptojacking which we will be discussed at length in the rest of the paper. 

\BfPara{Software-based Cryptojacking} \label{sec:sfcj}
Software-based cryptojacking involves installing a compromised binary on the target host that sends PoW solutions to a {\em dropzone} server. The most popular cryptojacking software is XMRig~\cite{ZimbaWMO20}, a cross-platform mining software supporting four different PoW protocols. Typically, XMRig is legitimately used by mining pools. However, its malware versions are also available and target non-miners. An XMRig-based malware called ``WaterMiner'' targets the online gaming community~\cite{ZimbaWMO20}).   

\BfPara{In-Browser Cryptojacking} \label{sec:brcj}
In-browser cryptojacking is done by injecting a {\em JavaScript} code in a website, allowing it to hijack the processing power of a visitor's device to mine a specific cryptocurrency. The precise nature of the cryptocurrency (\ie, mining protocol, difficulty, message exchange, \etc) is specified by the mining script embedded within a website. Upon visiting a website with cryptojacking code, the browser loads the webpage and executes the {\em JavaScript} snippet that contains instructions for mining and data transfer. As a result, the visiting host starts the mining activity by becoming part of a cryptojacking mining pool. A key feature of in-browser cryptojacking is being platform-independent: it can be run on any host, PC, mobile phone, tablet, etc., as long as the web browser running on this host supports {\em JavaScript}. {\em JavaScript} is one of the most popular web languages and, by default, is enabled in most major browsers. Furthermore, in-browser cryptojacking allows for mining at scale without requiring any custom hardware: as more visitors visit the website with cryptojacking scripts, more processing power is available. 

\subsubsection{Cryptojacking as a Replacement to Advertisement} \label{sec:adcj}
An ongoing debate sparked in the community for whether cryptojacking can serve as a replacement for online advertisement. Those advocating the approach have pointed out that users providing their CPU power to a website for mining can use the website without viewing online advertisements. Towards that, some websites, including the aforementioned `The Pirate Bay'', started using cryptojacking as a revenue substitute for online advertisements~\cite{Shaikh_17,Ernesto_17,Jones_2017} and become ``ads-free operation''. However, a counterargument to this model is the claimed to be the excessive abuse of the cryptojacking website to the visitor's CPU resources. In-browser cryptojacking scripts will not only run in the background without the user's consent. Still, they will also drain batteries in battery-powered platforms, indirectly affecting the user experience by locking the CPU power and not allowing him to use other applications. 

\section{Dataset and Preliminary Analysis}\label{sec:prelim}

\subsection{Data Collection} \label{sec:data}
We assembled a data set of cryptojacking websites published by Pixalate~\cite{Loechner17} and Netlab 360~\cite{Netlab360}. Pixalate is a network analytics company that provides data solutions for digital advertising. In Nov. 2017, they collected
a list of 5,000 cryptojacking websites actively stealing visitors’ processing power to mine cryptocurrency. We obtained a list cryptojacking websites from Pixalate. Netlab 360 (Network Security Research Lab at 360) is a data research platform that provides many datasets. From Netlab 360, we obtained  700 cryptojacking websites, released on Feb 24, 2018.

The top-level domain (TLD) distribution of the combined dataset, including the TLD type and the corresponding  percentage, is shown in~\autoref{tab:pixalate}. Unsurprisingly, .com and .net occupy the first and second spots of the top 10 TLDs represented in the dataset, with a combined total of 40.3\% of the websites belonging to them. Country-level domains have a significant presence, with countries such as Slovenia, Russia, and Brazil well represented in the dataset. New-gTLDs were also present in the top-10 gTLDs, with .site having $\approx$2.0\% of the sites. In Pixalate's dataset, six websites were found in the Alexa top 5000 websites, and 13 were among the Alexa top 10000 websites. Among the cryptojacking site, 68.3\% did not have a privacy policy.

In contrast, 56.8\% of websites had no ``terms and conditions'' statement, and  49.3\% did not have both a privacy policy and terms and conditions. This indicates that the majority of those websites could not formally, through those statements, inform their visitors of the usage of their  resources for mining cryptocurrencies, where cryptojacking is used instead of online advertisement \cite{CalzavaraRB18}. During our analysis, we  observed that 11\% of the websites in our dataset had stopped cryptojacking due to key revocation by the server, removal of the code from the website, or the closure of websites. We exclude them from our analysis. 

As mentioned in the~\tsref{sec:introduction}, among 5703 websites, we found 620 websites that are still active. Therefore, we revised our TLD distribution analysis in~\autoref{tab:pixalate} and reported our new findings in~\autoref{tab:newpixalate}. Our results show that among the active sites, (1) .com is still the most dominant TLD, (2) .sk has replaced .net at the second rank, and (3) country-level domains are now more prevalent in the dataset. Intuitively, this indicates that the attackers now focus on country-specific websites for cryptojacking attacks.

\begin{table}[t]
\centering
\begin{minipage}{.5\columnwidth}
\caption{Distribution of cryptojacking websites with respect to top-level domains in our dataset (type: generic, country, and new). }
\label{tab:pixalate}\vspace{-3mm}
\scalebox{0.68}{
\begin{tabular}{c|l|c|r|r}
\hline
{\textbf{Rank}} & {\textbf{TLD}} & {\textbf{Type}} & {\textbf{Sites}} & {\textbf{\%}} \\ \hline
1 & .com& g& 1945& 34.1\\ \hline
2 & .net & g &359& 6.2\\ \hline
3 & .si & c& 358& 6.2\\ \hline
4 & .online& g & 349 & 6.1\\ \hline
5 & .ru & c&242& 4.2\\ \hline
6 & .org& g&191& 3.3\\ \hline
7 & .sk& c&169& 2.9\\ \hline
8 & .info& g&169& 2.9\\ \hline
9 & .br& c &157& 2.7\\ \hline
10 & .site& n&116& 2.0\\ \hline
11 & others& {---}& 1648& 28.8\\ \hline 
\textbf{Total}  &   {---}& {---}  & \textbf{5703}& \textbf{100}\\ \hline
\end{tabular}}
\end{minipage}~
\begin{minipage}{.5\columnwidth}
\caption{Distribution of currently active cryptojacking websites with respect to top-level domains (type: generic, country, and new). }
\label{tab:newpixalate}\vspace{-3mm}
\scalebox{0.68}{
\begin{tabular}{c|l|c|r|r}
\hline
{\textbf{Rank}} & {\textbf{TLD}} & {\textbf{Type}} & {\textbf{Sites}} & {\textbf{Sites\%}} \\ \hline
1 & .com & g& 211& 34.0\\\hline
2 & .sk& c&145& 23.3\\ \hline
3 & .ru& c& 34& 5.5\\ \hline
4 & .pl& c&26& 4.2\\ \hline
5 & .net& g&25& 4.0\\ \hline
6 & .org & g&11& 1.8\\ \hline
7 & .ro& c&10& 1.6\\ \hline
8 & .de& c&10& 1.6\\ \hline
9 & .info& g&10& 1.6\\ \hline
10 & .id& c&8& 1.3\\ \hline
11 & others& {---}& 130& 20.9\\ \hline 
\textbf{Total}  &   {---}& {---}  & \textbf{620}& \textbf{100\%}\\ \hline
\end{tabular}}
\end{minipage}
\end{table}

\subsection{Methodology} \label{sec:method}
We perform static and dynamic analysis of the cryptojacking~{\em JavaScript} code. In the static analysis, we categorize the websites based on content and the currency they mine during cryptojacking. We extract the cryptojacking code and develop code-based features to examine their properties. We compare them, using those static properties, with malicious and benign {\em JavaScript} code. We use standard code analyzers to extract program-specific features. 

In our dynamic analysis, we explore the CPU power consumed by cryptojacking websites and its effects on user devices. We run test websites to mimic cryptojacking websites and carry out a series of experiments to validate our hypothesis. For our experiments, we use Selenium-based scripts to automate browsers and various end host devices, including Windows and Linux-operated laptops and an Android phone, to monitor the effect of cryptojacking under various operating systems and hardware architectures.  For website information, we use services provided by Alexa and SimilarWeb to extract information regarding website ranking, the volume of traffic, and the average time spent by visitors on that websites~\cite {Similarweb-18}. 

\section{Static Analysis} \label{sec:static}
For static analysis, we pursue three directions:  content-\, currency-, and code-based analysis. Content-based categorization provides insights into the nature of websites used for cryptojacking activities. In contrast, the currency-based categorization shows the distribution of service providers and platforms providing cryptojacking templates for those websites. The code-based analysis provides insight into the complexity of the cryptojacking scripts using various code complexity measures. Using those features, we perform two experiments for cryptojacking detection. A first experiment is a website-agnostic approach to uniquely distinguish cryptojacking {\em JavaScript} from other forms of malicious and benign {\em JavaScript} codes. A second experiment is a website-specific approach using which we detect 620 cryptojacking websites from non-cryptojacking websites.

\subsection{Content-based Categorization} \label{sec:contentcat} 
\begin{figure}[t]
\pgfplotsset{width=4cm,compat=1.5}  
       \centering
    \begin{tikzpicture}  
          
        \begin{axis}  
        [  
            ybar, 
            enlargelimits=0.07,  
            ylabel={Percentage}, 
            symbolic x coords={Business, Education, Adult, Entertainment, Illegal Content, Media, Tech Info, Shopping, Sport, Uncategorized}, 
            xtick=data,  
            x=0.7cm,
            bar width=2.8mm,
            legend style={at={(0.3,1.05)},anchor=west, legend columns=2},
            xticklabel style={rotate=30,anchor=east, align=right},
            grid=both,grid style={line width=.1pt, draw=gray!10},major grid style={line width=.1pt,draw=black!10}
            ]
        \addplot[color=black,pattern=crosshatch]  coordinates {(Business,21) (Education,19) (Adult,12) (Entertainment,9) (Illegal Content,6) (Media,2) (Tech Info,5) (Shopping,2) (Sport,1) (Uncategorized,23)};  
         \addlegendentry{Complete};
        \addplot[color=black, pattern=north west lines] coordinates {(Business,24) (Education,24) (Adult,3) (Entertainment,12) (Illegal Content,6) (Media,4) (Tech Info,3) (Shopping,5) (Sport,2) (Uncategorized,17)};  
        \addlegendentry{New};

        \end{axis}  
        \end{tikzpicture}  \vspace{-5mm}
\caption{The distribution of the two used datasets across various categories. The first figure is the complete dataset  of 5,703 websites, while the second is the currently active 620 websites (new). }
\label{fig:content}\vspace{-3mm}
    \end{figure}
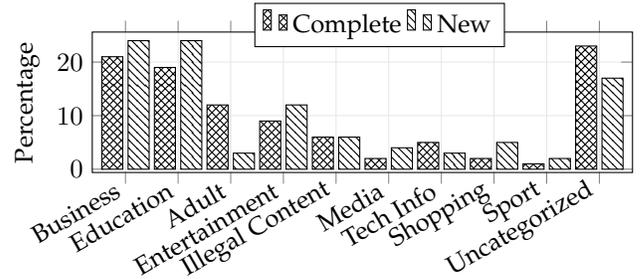

For a deeper insight into their usage, it is important to understand what kind of websites have cryptojacking scripts. To this end, and as a first step, we categorized the websites based on their contents into various categories using the {\em WebShrinker} website URL categorization API. {\em WebShrinker} assigns categories to websites based on the main usage of those websites using their contents. The results are shown in \autoref{fig:content}. We note that some websites are categorized as ``Illegal Content.'' These websites are primarily torrent websites that serve illegal copies of movies and software. Moreover, 19\% of websites were categorized as ``Education'', which can be attributed to the exploitation of trust by adversaries behind cryptojacking since educational sites are highly trusted~\cite{ZarrasKSHKV14}. \autoref{fig:content} shows the content-based categorization of 5703 websites in our complete dataset. Among them, 620 currently active websites have a different distribution, which we show in~\autoref{fig:content}. Among the currently active sites, 24\% are educational sites, and 24\% are business sites. This shows that attackers are still exploiting the trust associated with educational sites. 

\subsection{Currency-based Categorization} To understand the cryptojacking ecosystem, it is critical to find out what cryptocurrencies are typically being mined through in-browser cryptojacking. Therefore, we inspected the websites' scripts to extract information about the platforms and cryptocurrencies.
From our dataset, we found that there were eight platforms providing templates to mine two types of cryptocurrencies namely, Monero and JSEcoin. In~\autoref{tab:currencies}, we provide details about the eight platforms and their respective mining cryptocurrency. As a result, we found that a very large proportion of the websites ($\approx$81.57\%) use {\em Coinhive}~\cite{coinhive} platform to mine Monero cryptocurrency~\cite{Monero}, which is one of the few cryptocurrencies that supports in-browser mining. We found that $\approx$86.37\% of the websites in our dataset are mining Monero cryptocurrency through seven platforms. In addition, $\approx$2.61\% of the websites are using the JSEcoin platform~\cite{JSEcoin}, which is responsible for mining the JSEcoin cryptocurrency. 

Although PoW-based cryptocurrencies have many traits in common, they may vary in terms of their market cap, user base, application protocols, and mining rewards. In our dataset, we found two cryptocurrencies, namely Monero and JSEcoin, which are used for in-browser cryptojacking. In \autoref{tab:cct}, we report the differences between the two cryptocurrencies. While both of them are used for cryptojacking, 
at the time of writing this paper, JSEcoin was not launched in the market and did not have any ``Initial Coin Offering'' (ICO), which explains its low prevalence in our dataset. Furthermore, unlike Monero, which is resource-intensive, JSEcoin uses minimal CPU power and does not add a significant processing overhead to the target device. One of the key objectives of this paper is to characterize resource abuse in cryptocurrency mining, where Monero is shown to be a better example than the ``browser-friendly'' JSEcoin. Therefore, due to its high prevalence in the dataset, and the significant contribution towards the broader goal of this study, we mainly focus our work on Monero cryptocurrency. 

\begin{table}[t]
\small
\begin{center}
\caption{Detailed results of the currency-based analysis.  {$^1$ The variable name is abbreviated. No CJ: No cryptojacking. }}
\label{tab:currencies}\vspace{-3mm}
\scalebox{0.9}{
\begin{tabular}{lrr|crr}
\toprule
\multirow{2}{*}{Platform} & \multicolumn{2}{c|} {Websites} & \multirow{2}{*}{\Cc} & \multicolumn{2}{c} {Websites} \\
                           & \# & \% & & \# & \% \\
                        
\midrule
Coinhive    & 4652 & 81.57 & \multirow{7}{*}{Monero} & \multirow{7}{*}{4926}  & \multirow{7}{*}{86.37} \\ 
Hashing     & 67 & 1.17 &                          &                          &   \\ 
deepMiner   & 56 & 0.98 &                          &                          &   \\ 
Freecontent & 39 & 0.68 &                          &                          &   \\ 
Cryptoloot  & 38 & 0.67 &                          &                          &   \\ 
Miner       & 38 & 0.67  &                         &                          &   \\ 
Authedmine  & 35 & 0.61 &                          &                          &   \\ \hline
JSEcoin     & 149 & 2.61  & JSEcoin                & 149                      &  2.61 \\ \hline
No CJ   & 628 & 11.01  &              ---      & 628                       & 11.01      \\ \hline
Total       & 5703 & 100.00  &              ---      & 5703                       & 100.00      \\ 
\bottomrule

\end{tabular}}
\end{center}
\vspace{-3mm}\vspace{-3mm}
\end{table}

\subsection{Code-based Analysis}
We perform static analysis on the cryptojacking scripts to analyze the performance and complexity of their code. Static analysis reveals code-specific features for insights into the flow of information upon code execution. For static analysis, we gathered cryptojacking scripts from all major cryptojacking service providers in our dataset: {\em Coinhive}, JSEcoin, Crypto-Loot, Hashing, deepMiner, Freecontent, Miner, and Authedmine. We observed that all the service providers had unique codes, specific to their own platforms.  In other words, the websites using {\em Coinhive}{}'s services had the same {\em JavaScript} code template across all of them. Therefore, $\approx$81.57\% of the websites in our dataset were using the same {\em JavaScript} template for cryptojacking.
Similarly, all the websites using JSEcoin used the same  template for their mining. However, each provider's code template differed, which led us to believe that each script had unique static features. With all of that in mind, we performed static analysis on the cryptojacking websites and compared the results with another standard {\em JavaScript} for a baseline comparison.  

\begin{table}[t]

\caption{Comparison of Moneroe and JSEcoin. JSEcoin has not been released in the market as yet. }
\label{tab:cct}\vspace{-3mm}
\scalebox{0.9}{
\begin{tabular}{l|l|l|c|l}
\hline
\textbf{Currency} & \textbf{\begin{tabular}[c]{@{}l@{}}Market \\ Cap\end{tabular}} & \textbf{\begin{tabular}[c]{@{}l@{}}Consensus\\ Algorithm\end{tabular}} & \textbf{\begin{tabular}[c]{@{}l@{}}Resource\\ Intensive\end{tabular}} & \textbf{\begin{tabular}[c]{@{}l@{}}Dataset\\ Prevelance\end{tabular}} \\ \hline
Monero            & 2.3B     & CryptoNight  &   \cmark & 86.37\%  \\ \hline
JSEcoin           & ---     & SHA-256    & \xmark  & 2.61\%   \\ \hline
\end{tabular}}\vspace{-3mm}\vspace{-3mm}
\end{table}

\begin{table*}[htb]
\centering
\caption{The static features of the cryptojacking, malicious, and benign samples. The mean ($\mu$) and standard deviation ($\sigma$) of the features are also reported.}
\vspace{-3mm}
\label{tab:fcj}
\scalebox{0.6}{
\begin{tabular}{|l|l|r|r|r|r|r|r|r|r|r|r|r|r|r|r|r|r|r|}
\hline
Cat.   & Platforms& $M$   & $M_d$   & $B$      & $D$       & $E$           & $c_l$   & $T$          & $\eta$   & $V$         &  $\eta _{1}$  & $n_1$   &  $\eta _{2}$ & $n_2$   & params & sloc & physical & $M_s$     \\ \hline

\multirow{10}{*}{\begin{turn}{-90} Cryptojacking \end{turn}} 
&deepMiner   & 184 & 44.2 & 14.1  & 113.0 & 4,810,434 & 4,667 & 267,246 & 554 & 42,533   & 47 & 2,440 & 507  & 2,227 & 75  & 416  & 499 & 67.8 \\ \cline{2-19}
&Authedmine  & 168 & 26.5 & 19.7  & 82.8  & 4,912,255 & 6,096 & 272,903   & 844 & 59,259  & 41 & 3,247   & 803 & 2,849  & 73  & 633  & 784 & 62.8 \\ \cline{2-19}
&Hashing     & 138 & 29.1 & 7.2   & 94.6  & 2,185,379 & 2,794 & 124,138 & 342 & 24,393   & 38 & 1,469 & 315  & 1,415 & 37  & 412  & 505 & 68.2 \\ \cline{2-19}
&Miner       & 133 & 27.7 & 9.3   & 90.5  & 2,537,930 & 3,239 & 140,996 & 403 & 28,032   & 39 & 1,690 & 364  & 1,549 & 49  & 479  & 617 & 64.1 \\ \cline{2-19}
&Coinhive    & 131 & 27.5 & 9.1   & 94.8  & 2,608,021   & 3,226 & 144,890    & 368 & 274,970      & 37 & 1,697   & 331 & 1,529  & 48  & 476  & 594 & 63.7 \\ \cline{2-19}
&Crypto-loot & 128 & 39.7 & 11.4  & 88.1  & 3,034,935 & 3,788 & 168,607 & 546 & 34,443   & 45 & 1,962 & 501  & 1,826 & 62  & 322  & 389 & 70.3 \\ \cline{2-19}
&Freecontent & 117 & 28.3 & 8.1   & 89.4  & 2,180,394 & 2,884 & 121,133   & 350 & 24,373   & 38 & 1,469 & 312  & 1,415 & 37  & 412  & 505 & 62.7 \\ \cline{2-19}
&JSEcoin     & 64  & 17.2 & 10.2 & 62.9 & 1,945,165   & 3,257 & 108,064 & 716 & 30,888   & 45 & 1,878 & 671  & 1,379 & 49  & 372  & 412 & 64.7 \\ \cline{2-19}
&\textbf{Mean} ($\mu$)	     &130.3	&	29.9	&	11.3	&	88.9	&	3,026,191	&	3,755.1	&	168,121	&	516.4	&	33,925	&	41.3	&	1,981.5	&	475.1	&	1,773.6	&	53.8	&	440.3	&	538.1	&	64.9 \\ \cline{2-19}
&\textbf{SD.} ($\sigma$)	&	35.9	&	8.4	&	3.9	&	13.8	&	1,180,403	&	1,109.9	&	65,577	&	185.1	&	11,856	&	3.9	&	599.3	&	182.8	&	519.3	&	14.8	&	93.2	&	126.3	&	2.8 \\
 \hline
\multirow{8}{*}{\begin{turn}{-90} Malicious \end{turn}} 
    & 20160209  & 92 & 21.5 & 5.6  & 25.1 & 423,925   & 1,833 & 23,551 & 580  & 16,826 & 27 &  1,032 & 553   & 801  & 22  & 427  & 503  & 44.4 \\ \cline{2-19}
    & 20161126  & 62 & 15.3 & 4.2  & 24.6 & 315,735   & 1,563 & 17,540 & 292  & 12,800 & 17 &  798 & 275     & 765  & 0   & 403  & 481  & 90.5 \\ \cline{2-19}
    & 20170110  & 14 & 4.4  & 15.0   & 26.7 & 1,211,305 & 4,704 & 67,294 & 782  & 45,210 & 15 &  2,740 & 767   & 1,964& 232 & 313  & 564  & 93.6 \\ \cline{2-19}
    & 20170507  & 6  & 24.0   & 5.9  & 11.1 & 199,917   & 1,864 & 11,106 & 777  & 17,897 & 18 &  942 & 759     & 922  & 1   & 25   & 890  & 71.7 \\ \cline{2-19}
    & 20160927  & 3  & 1.4  & 4.0   & 32.5 & 393,555   & 1,575 & 21,864 & 204  & 12,084   & 13 &  957 & 191     & 618  & 0   & 213  & 98   & 23.2 \\ \cline{2-19}
    & 20170322 & 2  & 18.1 & 11.8 & 7.1  & 253,442   & 3,514 & 14,080 & 1,123 & 35,607 & 9  &  1,762 & 1,114    & 1,752& 3   & 11   & 1,738& 90.9 \\ \cline{2-19}
    & 20170303  & 2  & 8.6  & 0.2  & 9.4  & 8,338     & 147   & 463    & 63   & 878    & 13 &   73 & 50      & 74   & 4   & 23   & 55   & 78.7 \\ \cline{2-19}
    & 20160407  & 1  & 33.3 & 0.1  & 2.7  & 207       & 19    & 11     & 16   & 76       & 5  &  12    & 11       & 7    & 0   & 3    & 3    & 78.9 \\ \cline{2-19}
    & 20170501  & 1  & 0.9  & 2.1  & 3.3  & 21,464     & 758   & 1,192   & 322  & 6,314   & 5  &    431 & 317      & 327  & 0   & 105  & 105  & 35.9 \\ \cline{2-19}
    & 20160810  & 1  & 12.5 & 0.5  & 11.9 & 20,148    & 275   & 1,119  & 70   & 1,685  & 6  &    255 & 64      & 20   & 0   & 8    & 13   & 60.4 \\ \cline{2-19}
&\textbf{Mean} ($\mu$)	&	18.4	&	14	&	4.9	&	15.5	&	284,803.7	&	1,625.2	&	15,822	&	422.9	&	14,938	&	12.8	&	900.2	&	410.1	&	725	&	26.2	&	153.1	&	445	&	66.9\\ \cline{2-19}
&\textbf{SD.} ($\sigma$)	&	31.9	&	10.5	&	5	&	10.8	&	364,470.8	&	1,508.9	&	20,248	&	374.8	&	15,045	&	6.9	&	834.7	&	372.5	&	686.6	&	72.6	&	171.9	&	543.5	&	24.9\\  
\hline
\multirow{8}{*}{\begin{turn}{-90} Benign \end{turn}} 
    & The Boat        & 2,135 & 69.3  & 110.8 & 392.0   & 130,285,522 & 31,916 & 7,238,084   & 1,364 & 332,361   & 59 &  17,341    & 1,305   & 14,575  & 852 & 3,084 & 3,349 & 66.7 \\ \cline{2-19}
    & IBM Design      & 2,119 & 68.3  & 110.9 & 397.1 & 132,237,213 & 32,018 & 7,346,511   & 1,351 & 332,981   & 59 &   17,393   &    1,292 & 1,4625 & 853 & 3,103 & 3,372 & 66.7 \\ \cline{2-19}
    & Histography     & 1,743  & 40.7  & 95.2  & 249.5 & 71,325,242    & 26,627  & 3,962,513     & 1,704 & 285,833    & 55 &  14,963  & 1,649    & 11,663     & 803 & 4,278  & 5,043  & 59.4 \\ \cline{2-19}
    & Know Lupus      & 1,006 & 28.1  & 92.9  & 170.4 & 47,474,425  & 25,120 & 2,637,468   & 2,181 & 278,600    & 54 &  13,424    & 2,127   & 11,696  & 615 & 3,583 & 4,288 & 65.2 \\ \cline{2-19}
    & tota11y         & 815   & 38.8  & 59.4  & 227.7 & 40,563,065  & 17,486 & 2,253,503   & 1,167 & 178,157   & 52 &   9,764   & 1,115    & 7,722  & 412 & 2,099 & 2,336 & 62.9 \\ \cline{2-19}
    & Masi Tupungato  & 784   & 58.2  & 47.1  & 185.0   & 26,199,193  & 14,296 & 1,455,510 & 958  & 141,585   & 43 &  7,875    & 915    & 6,421  & 238 & 1,347 & 1,470 & 67.2 \\ \cline{2-19}
    & Fillipo         & 703   & 42.9  & 43.1  & 194.3 & 25,139,766  & 12,900 & 1,396,653   & 1,045 & 129,377   & 54 &  7,132      &   991   & 5,768 & 269 & 1,637 & 1,770 & 61.5 \\ \cline{2-19}
    & Leg Work        & 412   & 75.7  & 34.0    & 241.3 & 24,651,056  & 11,100 & 1,369,503   & 589  & 102,143   & 45 &     5,835   & 544    & 5,265  & 66  & 544   & 633   & 65.9 \\ \cline{2-19}
    & Code Conf       & 409   & 27.8  & 41.1  & 197.1 & 24,336,420  & 12,500 & 1,352,023   & 939  & 123,437   & 49 &  7,162      &    890  & 5,338  & 315 & 1,469 & 1,753 & 64.9 \\ \cline{2-19}
    & Louis Browns    & 368   & 35.6  & 21.2  & 106.7 & 6,792,400   & 6,529  & 377,355     & 862  & 63,667    & 51 &   3,393     &    811  & 3,136  & 68  & 1,034 & 1,357 & 53.3 \\ \cline{2-19}
    & \textbf{Mean} ($\mu$)	&	1,049.4	&	48.5	&	65.6	&	236.1	&	52,900,430	&	19,049.2	&	2,938,912	&	1,216	&	196,814	&	52.1	&	10,428.2	&	1,163.9	&	8,621	&	449.1	&	2,217.8	&	2,537.1	&	63.4\\ \cline{2-19}
& \textbf{SD.} ($\sigma$)	&	694	&	17.8	&	33.6	&	92.8	&	44,755,377	&	9,151.2	&	2,486,409	&	459.8	&	100,856	&	5.3	&	4,999	&	456.7	&	4,165	&	310.3	&	1,225.4	&	1,418.2	&	4.3\\ 
\hline
\end{tabular}}
\vspace{-3mm}\vspace{-3mm}
\end{table*}

\subsubsection{Data Attributes} \label{sec:dc}
We prepared our dataset for static analysis by collecting all of the popular cryptojacking scripts from our list of websites. We found eight unique scripts among all the websites, each belonging to one service provider. As a control experiment, we collected an equal number of malicious and benign {\em JavaScript} codes to design machine learning models for detection. We aimed to obtain a set of features unique only to the cryptojacking scripts and aid in their detection. With such knowledge of those features, more accurate countermeasures can be further developed that will accurately predict if a given host machine is under a cryptojacking attack. 
To avoid bias towards a certain class, we were limited to including equal sizes of malicious and benign {\em JavaScript} samples for the static analysis. Although there are many samples of malicious and benign {\em JavaScript} in the wild, only eight cryptojacking scripts are available in comparison. Since our work is focused on distinguishing cryptojacking scripts from malicious and benign {\em JavaScript}, we had to balance the size of each class. While the number of scripts might seem like a limitation of our work, we believe the promise of this work is substantial. As more currencies and platforms use cryptojacking, more samples will be available for a broader study. Demonstrating a baseline analysis to support the argument that cryptojacking scripts are uniquely identifiable can open further analysis of cryptojacking scripts across well-understood analysis tools, which we explore in this paper.

In lieu, we used the existing data of the cryptojacking websites (\textsection\ref{sec:data}) and online resources from GitHub for malicious {\em JavaScript} sample \cite{Wizche17,Petrak17} . For benign {\em JavaScript}, we used the set of non-cryptojacking websites and parsed their HTML code to extract benign {\em JavaScript} code~\cite{staff_2017}. In summary, we had 8 samples of cryptojacking~{\em JavaScript}, spanning all the websites. Accordingly, we selected 10 malicious and 10 benign scripts for our machine-learning model.

\subsubsection{Feature Extraction} \label{sec:feat}
We describe the features we used for our static analysis of cryptojacking, malicious, and benign codes in the following.

\BfPara{Cyclomatic Complexity} Cyclomatic complexity~\cite{WatsonMW96} measures the complexity of code  using a control flow graph (CFG), where each node represents a function and a directed edge between two nodes indicates a caller-callee relationship. Let $E$ be the number of edges, $N$ be the number of nodes, and $Q$ be the number of connected components in the CFG, $M$ can be used to denote the cyclomatic complexity of the program and is calculated as $M = E + 2Q - N$.

\BfPara{Cyclomatic Complexity Density} Cyclomatic complexity density~\cite{FentonN99} measures Cyclomatic complexity, defined above, spread over the total code length. Usually, malware authors obfuscate their code to avoid detection \cite{Galeano17,MohaisenAM15,KangJMK15,AlasmaryKAPCAAN19,MohaisenA13}. As such, they may alter the flow of a program and add extra functions. While adding more functions and lines of code will undoubtedly increase the size of the code, its complexity will remain the same, which could be used as a feature of their detection. Let $c_l$ be the total number of lines of code, then the cyclomatic complexity density, denoted by $M_d$, can be computed as $M_d = \frac{E + 2Q - N}{c_l}$

\BfPara{Halstead Complexity Measures} The Halstead complexity measures are used as metrics to characterize the algorithmic implementation of a programming language~\cite{Serebrenik11}. Those measures include the vocabulary $\eta$, the program length $n$, the calculated program length $n_c$, the volume $V$, the effort $E$, the delivered bugs $B$, the time $T$, and the difficulty $D$. Let the number of distinct operators be $\eta _{1}$, the number of distinct operands be $\eta _{2}$, the total number of operators be $n_1$, the total number of operands be $n_2$, the $\eta, n, n_l, V, E,$ and $B$ are defined as follows:
\begin{align}\label{equation:halsted} 
    \nonumber\eta &= \eta _{1}+ \eta _{2},  & n &= n_{1}+n_{2}\\
    \nonumber n_c &= (\eta _{1}\log _{2}\eta _{1})+(\eta _{2}\log _{2}\eta _{2}), & V &= n\times \log _{2}\eta\\
    \nonumber D &= (\eta _{1}/ 2)\times (n_{2} / \eta _{2}), & E &= D\times V\\
    \nonumber T &= {(D \times V)}/{18}, & B &= {E^{\frac{2}{3}}}/{3000} 
\end{align}

\BfPara{Maintainability Score} The maintainability score $M_s$ is calculated using Halstead volume $V$, cyclomatic complexity $M$, and the total lines of code in the {\em JavaScript} file $c_l$. The maintainability score index $M_i$ is calculated between [0-100] and is defined as $M_s  =  171 - 5.2 \log (V) - 0.23M  - 16.2 \log (c_l); M_i = \max(0, \frac{M_s}{171})$.

\BfPara{Source Lines of Code} Source lines of code (SLOC) measure the lines of code in the program after excluding the white spaces. SLOC is a predictive parameter to evaluate the effort required to execute the program. It also provides insights into program maintainability and productivity.

\BfPara{\em Results} To extract features in our code-based analysis, we used \pl, a {\em JavaScript} static analysis and source code complexity tool~\cite{badge_2016}. For each {\em JavaScript} code, we ran \pl and record the 17 extracted features as reported in~\autoref{tab:fcj}. From~\autoref{tab:fcj}, we observed that certain features, such as $M$, $M_d$, $V$, and $T$, are clearly discriminative among all the categories. For further analysis, in the next section, we will look into the correlation of these features among each category to see whether there is a unique pattern among each category, which allow us to build a machine learning system that can automatically identify different {\em JavaScript} categories based on the extracted features.

\subsubsection{Correlation Analysis} \label{sec:correlation}
While meaningful, the individual features among those analyzed above might not shed light on their distinguishing power, given their large numbers. To this end, we  pursue a correlation analysis to understand their patterns. In particular, we conducted a correlation analysis to observe the similarity of features among the three categories of scripts, cryptojacking, malicious, and benign. The correlation analysis showed the consistency of the relationship distinctive to each category of the {\em JavaScript} codes. As such, this gave us insights into coding patterns and features unique to the style of coding cryptojacking, malware, and benign scripts. We computed the correlation of the features in all the scripts belonging to each category of {\em JavaScript}. We used the Pearson correlation coefficient for this analysis, which is defined as $\rho(X,Y) = {{Cov}(X,Y)}/({\sqrt{{Var}(X){Var}(Y)}})$, where $X$ and $Y$ are the random variables, $Var$ and $Cov$ are the {\em variance} and {\em covariance} of the random variables, respectively. 

We performed a comparative analysis on the correlation matrix obtained for each class to identify distinguishing features and reasons for their prevalence in cryptojacking~{\em JavaScript}. In \autoref{algo:corr}, we outline the procedure for identifying those features. The algorithm takes as an input the correlation matrix of cryptojacking \textbf{C}, malicious \textbf{M}, and benign \textbf{B} {\em JavaScript} features reported in \autoref{tab:fcj}, computes the mean of the column vector concerning one feature in the row, compares the mean feature of each class, and outputs the most distinguishing features in cryptojacking scripts that are highly correlated within their class. The distinguishing aspect of a feature in cryptojacking class is obtained by subtracting its mean value from complementary mean values of features from the other two classes and selecting the maximum difference.

\begin{algorithm}[t]  

\SetKwInOut{Input}{Inputs}  
\Input{\textbf{C}, \textbf{M}, \textbf{B}, $i = |\textbf{C}|$;\\
}

\textbf{Initialize Lists} $C_{m}, M_{m}, B_{m}, Arr$ = []\;

\For{$k = 0;\ k < i;\ k = k + 1$}{
$C_{m}[k] \leftarrow \tfrac{(\sum_{j=1}^{{i}} {c(i,j)}} {{i}}$, 
$M_{m}[k]\leftarrow \tfrac{(\sum_{j=1}^{{i}} {m(i,j)}} {{i}}$,
$B_{m}[k]\leftarrow\tfrac{(\sum_{j=1}^{{i}} {c(i,j)}} {{i}}$\;
\If{$\left(\left(C_{m}[k] - M_{m}[k]\right) \textbf{and} \left(C_{m}[k] - B_{m}[k]\right)\right) >  (M_{m}[k] - B_{m}[k])$ }{
$Arr \leftarrow C_{m}[k]$\;}}
\SetKwInput{KwData}{Output}
 \KwData{$Arr$ }

  \caption{Identifying Significant Features}
\label{algo:corr}
\end{algorithm}

\begin{figure*}[!t]
\begin{subfigure}[Correlation in cryptojacking JavaScript
\label{fig:crypto}]{\includegraphics[width=0.32\textwidth]{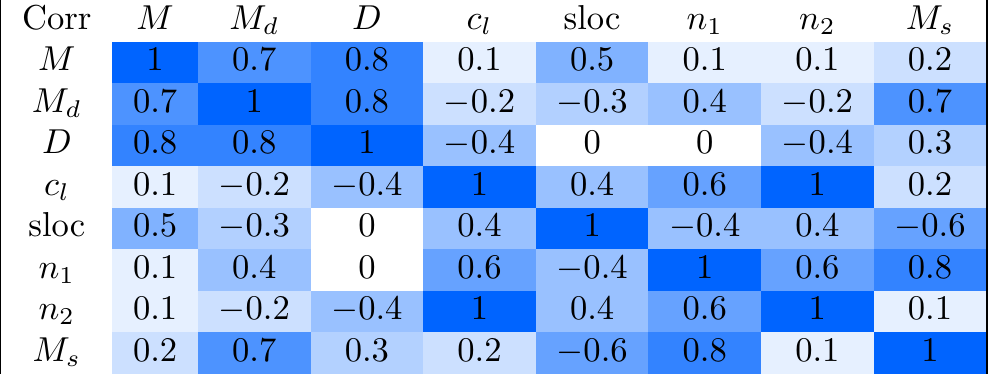}} 
~
\end{subfigure}
\begin{subfigure}[Correlation in malicious JavaScript \label{fig:maliciousy}]{\includegraphics[width=0.32\textwidth]{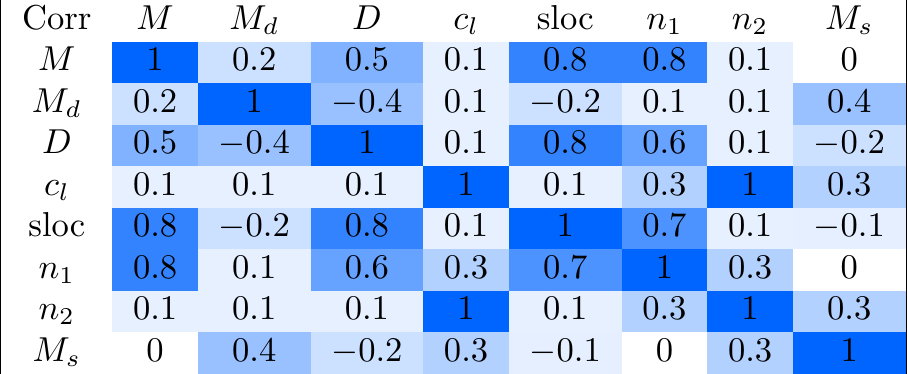}}
~
\end{subfigure}
\begin{subfigure}[Correlation in benign JavaScript \label{fig:benign}]{\includegraphics[width=0.32\textwidth]{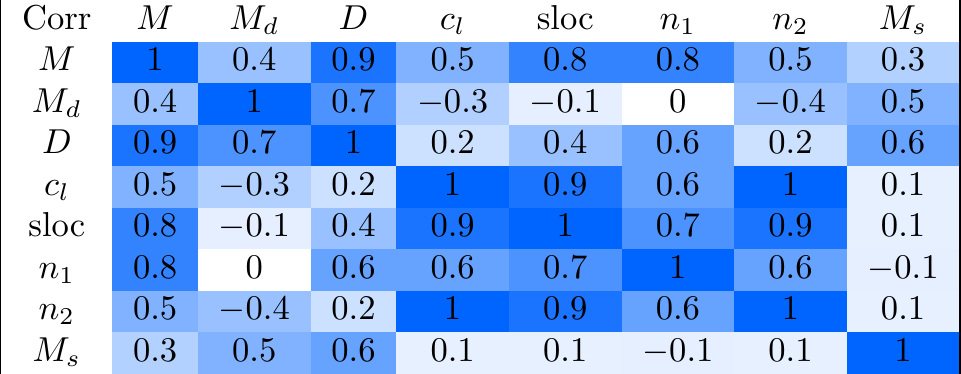}} 
\end{subfigure}
\vspace{-3mm}\vspace{-2mm}
\caption{Heatmap of correlation coefficients among the features of three categories of JavaScript. These are the subset features of \autoref{tab:fcj}, obtained using \autoref{algo:corr}. It can be noted that features among benign scripts appear to be highly correlated, while the features among malicious scripts remain highly uncorrelated. Correlation among the features of cryptojacking scripts remains in the middle, relative to the other two.    } 
\label{fig:correlation}\vspace{-3mm}\vspace{-3mm}
\end{figure*}

The output $Array$, in \autoref{algo:corr}, contains a subset of features from the seventeen features unique to cryptojacking scripts. In particular, we found eight features and plotted their result in~\autoref{fig:correlation}. It can be observed that cryptojacking scripts are more correlated concerning the cyclomatic complexity density $M_d$ and the maintainability score $M_s$. In contrast, malicious and benign scripts are not as correlated over those same parameters. From the description of those features provided in \textsection\ref{sec:feat}, more profound insights can be developed regarding the coding patterns, code complexity, CFGs, and maintainability of cryptojacking scripts. Furthermore, high correlation provides insights into code contents: all cryptojacking scripts must perform a sequence of similar actions with complementary execution patterns and information flows. We apply this understanding in our dynamic analysis and validate it using WebSocket inspection. 

\subsection{Classification Models} \label{sec:clustering}
After collecting samples from the three different classes of {\em JavaScript} codes, we used a machine-learning approach for their classification. Our objective was to construct a model capable of learning various code-based features of malicious, benign, and cryptojacking code samples and use them as a classification primitive for accurate detection. To that end, we used three well-known machine learning techniques, namely Logistic Regression (LR), Linear Discriminant Analysis (LDA), k-nearest neighbors (k-NN), Support Vector Machines (SVM), and Random Forest (RF)~\cite{LiuCY09}. We apply these techniques in two experiments. In the first experiment, we detect eight unique cryptojacking platforms from malicious and benign {\em JavaScript}. The second experiment presents a website-specific detection approach that uniquely distinguishes between 620 cryptojacking and non-crypto-jacking websites. In the following, we provide a brief background of our detection techniques.

\BfPara{Evaluation and Results}  We conduct two experiments to evaluate the static analysis. For both experiments, we used scikit-learn (in Python) to implement logistic regression, LDA, k-NN, SVM, and Random Forest classifiers. 

In the first experiment, we used the features in \autoref{tab:fcj} as input and assigned three distinct classes for the corresponding category. We performed each experiment 20 times and reported the average in \autoref{tab:confusionmatrix}. We report each model's precision, recall, and F1 score for evaluation. Our results show that logistic regression and LDA performed well, achieving an accuracy of 100\% as indicated by the value 1.00 for precision, recall, and F1-score. In contrast, k-NN performed relatively poorly with 0.93, 0.90, and 0.91 values for precision, recall, and F1-score. SVM and Random Forest performed better than k-NN with 0.96, 0.95, and 0.95 values for precision, recall, and F1-score. As a result, we derive two key conclusions from our experiments. First, the parametric evaluation models are more suitable for our classification problem and can serve well to detect cryptojacking scripts among other {\em JavaScript} codes. Second, the features of the three classes of {\em JavaScript} are highly discriminative, indicating unique coding patterns for each category, which are easily and accurately distinguishable.

For the second experiment,  we scanned all websites and identified 620 websites that were still up with the cryptojacking scripts in them. For each website, we extracted all {\em JavaScript} codes present on the website, including the cryptojacking scripts. We then randomly selected 620 benign websites from Alexa's top 1 Million that 1) did not have cryptojacking code and 2) were marked safe by VirusTotal from other forms of malicious {\em JavaScript}. From this dataset, we created two classes for static analysis. In the first class, we extracted all {\em JavaScript} features of 620 cryptojacking websites. In the second class, we extracted all {\em JavaScript} features of 620 non-cryptojacking websites. For each class, we extracted the same features reported in Table 4 (\ie cyclomatic complexity, Halstead difficulty, and distinct operands \etc). By examining the features manually, we observed that cryptojacking websites had highly discriminative features from non-cryptojacking websites, which invariably supported the discrimination power of our detection models in identifying the two classes. For independently reproducible results, we have released our dataset publicly~\cite{Saad20}. 

We divided our dataset into 75\% training and 25\% testing subsets. The results from the second experiment show that all classification models achieved an accuracy of 100\% as indicated by the value 1.00 for precision, recall, and F1-score in~\autoref{tab:confusionmatrix_two}. Our second experiment validates that the features of cryptojacking scripts are highly discriminative from other {\em JavaScript} codes, which can be easily detected across cryptojacking and non-cryptojacking websites.

\section{Dynamic Analysis} \label{sec:dynamic}

\begin{table}[t]
\begin{minipage}{.5\columnwidth}
\centering
\caption{Classification performance (complete dataset) against the F1-score, precision and recall.}\vspace{-3mm}
\scalebox{0.75}{
\begin{tabular}{llll}
\hline
                                                   & \textbf{F1}         & \textbf{Pre}        & \textbf{Rec}           \\ \hline
\multicolumn{1}{l|}{\textbf{LR}} & \multicolumn{1}{l|}{1.00} & \multicolumn{1}{l|}{1.00} & \multicolumn{1}{l}{1.00} \\ \hline
\multicolumn{1}{l|}{\textbf{LDA}} & \multicolumn{1}{l|}{1.00} & \multicolumn{1}{l|}{1.00} & \multicolumn{1}{l}{1.00} \\ \hline
\multicolumn{1}{l|}{\textbf{k-NN}} & \multicolumn{1}{l|}{0.91} & \multicolumn{1}{l|}{0.93} & \multicolumn{1}{l}{0.90} \\ \hline
\multicolumn{1}{l|}{\textbf{SVM}} & \multicolumn{1}{l|}{0.96} & \multicolumn{1}{l|}{0.95} & \multicolumn{1}{l}{0.95} \\ \hline
\multicolumn{1}{l|}{\textbf{RF}} & \multicolumn{1}{l|}{0.96} & \multicolumn{1}{l|}{0.95} & \multicolumn{1}{l}{0.95} \\ \hline
\end{tabular}}
\label{tab:confusionmatrix}  
\end{minipage}~
\begin{minipage}{.5\columnwidth}
\centering
\caption{Classification performance (new dataset) against the F1-score, precision and recall}\vspace{-3mm}
\scalebox{0.75}{
\begin{tabular}{llll}
\hline
                                                   & \textbf{F1}         & \textbf{Pre}        & \textbf{Rec}           \\ \hline
\multicolumn{1}{l|}{\textbf{LR}} & \multicolumn{1}{l|}{1.00} & \multicolumn{1}{l|}{1.00} & \multicolumn{1}{l}{1.00} \\ \hline
\multicolumn{1}{l|}{\textbf{LDA}} & \multicolumn{1}{l|}{1.00} & \multicolumn{1}{l|}{1.00} & \multicolumn{1}{l}{1.00} \\ \hline
\multicolumn{1}{l|}{\textbf{k-NN}} & \multicolumn{1}{l|}{1.00} & \multicolumn{1}{l|}{1.00} & \multicolumn{1}{l}{1.00} \\ \hline
\multicolumn{1}{l|}{\textbf{SVM}} & \multicolumn{1}{l|}{1.00} & \multicolumn{1}{l|}{1.00} & \multicolumn{1}{l}{1.00} \\ \hline
\multicolumn{1}{l|}{\textbf{RF}} & \multicolumn{1}{l|}{1.00} & \multicolumn{1}{l|}{1.00} & \multicolumn{1}{l}{1.00} \\ \hline
\end{tabular}}
\label{tab:confusionmatrix_two}  
\end{minipage}
\end{table}

\begin{figure}[t]
\begin{center}
    \includegraphics[ width=0.45\textwidth]{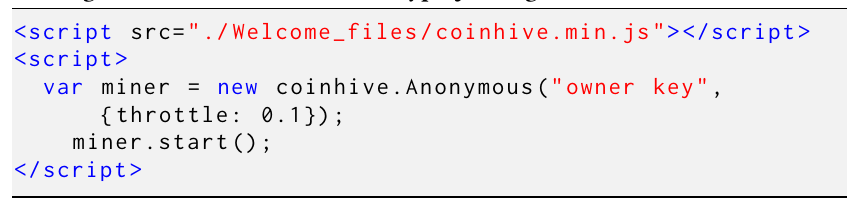}\vspace{-4mm}
        \caption{Malicious JavaScript code that links a website to Coinhive.}
\label{lst:coinhive}\vspace{-4mm}
\end{center}
\end{figure}

\subsection{Resource Consumption Profiling} \label{sec:processor}

\BfPara{Settings and Measurements Environment} 
We noticed that in each cryptojacking website, a {\em JavaScript} snippet encodes a key belonging to the code owner and a link to a server to  which the PoW is sent. \autoref{lst:coinhive} provides a script found in websites that use {\em Coinhive} for mining. The source (\textit{src}) refers to the actual {\em JavaScript} file that is executed after a browser loads the script. In this script, we also noticed a {\em throttling parameter}, which controls how many resources a cryptojacking script uses on the host. We use the throttling parameter, $\alpha$, as an additional variable in our experiment. We experiment with $\alpha=\{0.1, 0.5, 0.9\}$. 

\begin{figure}[t]
\begin{subfigure}[JavaScript enabled\label{fig:dyn}]{\includegraphics[width=0.24\textwidth]{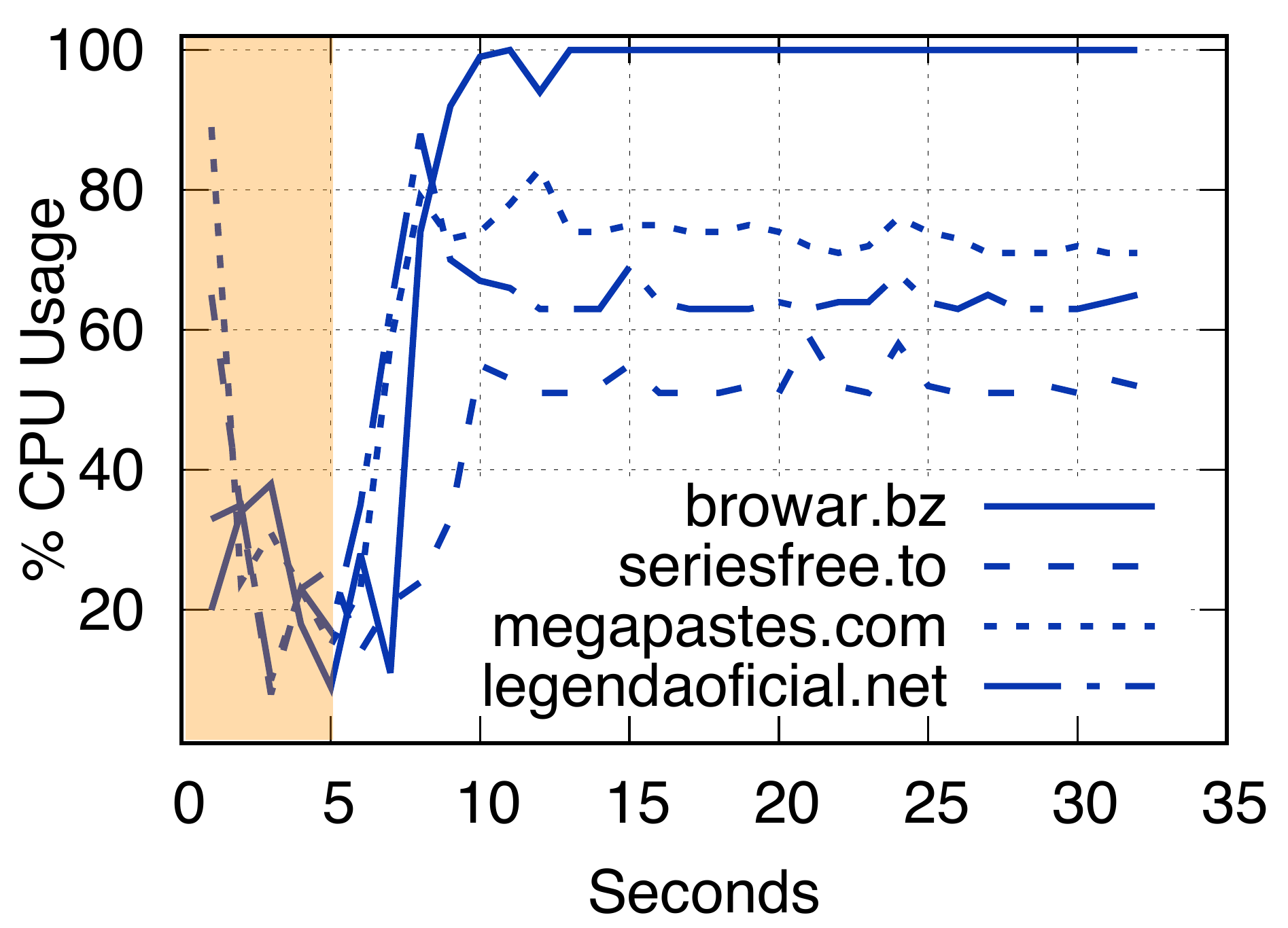}} 
\end{subfigure}
\begin{subfigure}[JavaScript disabled\label{fig:maliciousx}]{\includegraphics[width=0.24\textwidth]{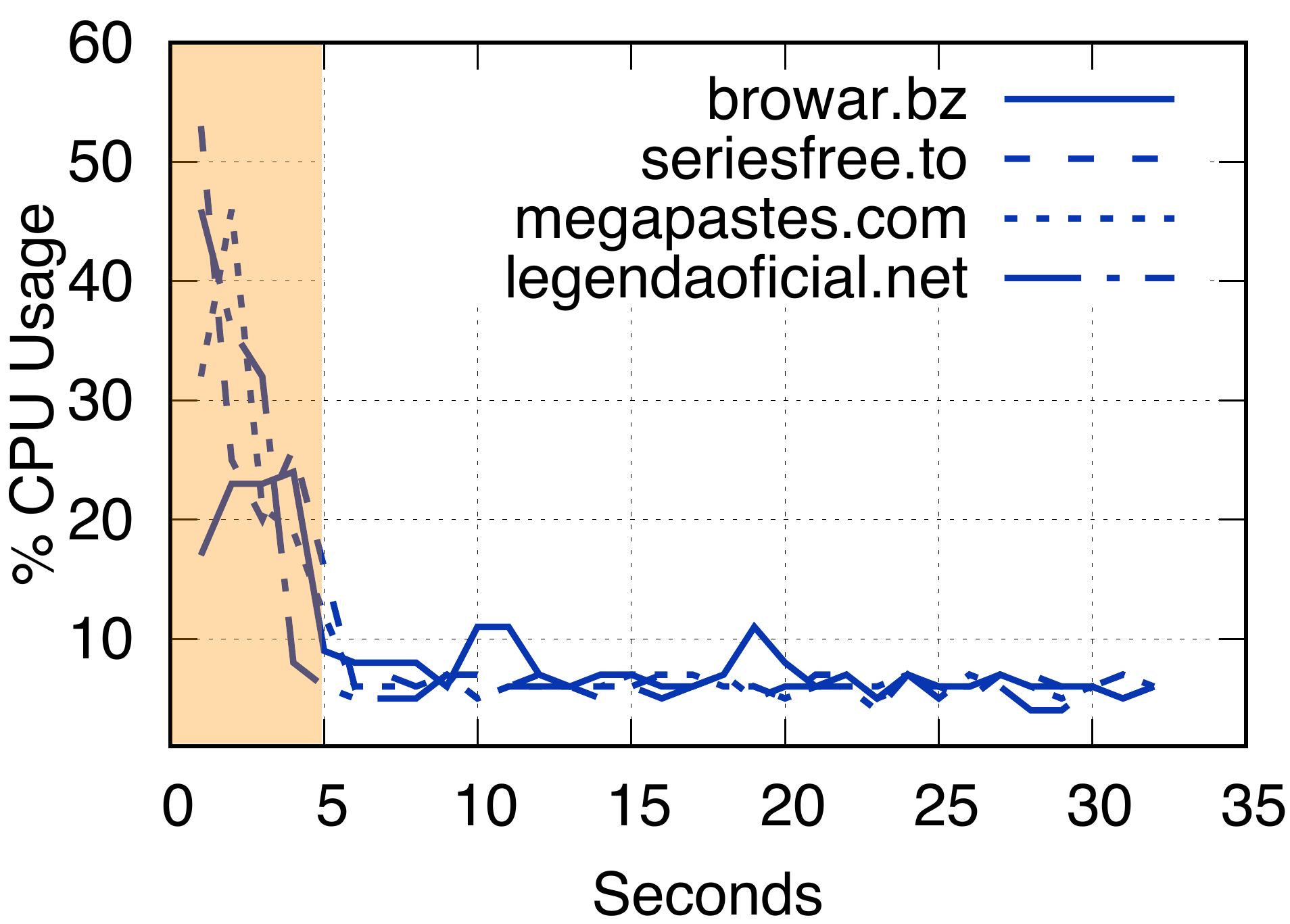}}
\end{subfigure}
\vspace{-5mm}
\caption{Processor usage by four cryptojacking websites with JavaScript enabled and disabled. }
\label{fig:dynamic}\vspace{-3mm}
\end{figure}

To understand the impact of cryptojacking on resource usage in different platforms, we use battery-powered machines running Microsoft Windows, Linux, and Android operating systems (OSes). We selected three laptops, each with one of those OSes. The Windows laptop was Asus V502U (Intel Core i7-6500U processor @ 3.16 GHz), the Linux laptop was Lenovo G50 (Intel Core i5-5200U, processor (4 cores) @ 2.20 GHz), and the Android phone was {\em Samsung Galaxy J5} with Android version of 6.0.1.

Using the above parameters, we set up an account on {\em Coinhive} for a a key that links our ``experiment website'' to the server. We embedded the code in~\autoref{lst:coinhive} in the website's HTML tags. To measure the usage of resources while running cryptojacking websites, we set up a Selenium-based web browser automation and ran cryptojacking websites for various evaluations. Selenium is a portable web-testing software miming actual web browsers~\cite{bruns2009web,seleniumdocumentation}.

\BfPara{CPU Usage} 
First, we baseline our study to highlight CPU usage as a fingerprint across multiple websites that employ cryptojacking using the aforementioned configurations and measurement environment. We study the usage of CPU with and without cryptojacking in place. For this experiment, we select four cryptojacking websites. To measure the impact of cryptojacking on CPU usage, we ran those websites in our \slx environment, for 30 seconds, with {\em JavaScript} enabled (thus running the cryptojacking scripts) and disabled (baseline; not running the cryptojacking scripts). We use this test experiment as our control.

\begin{figure*}[t]
\begin{centering}
\includegraphics[ width=0.9\textwidth]{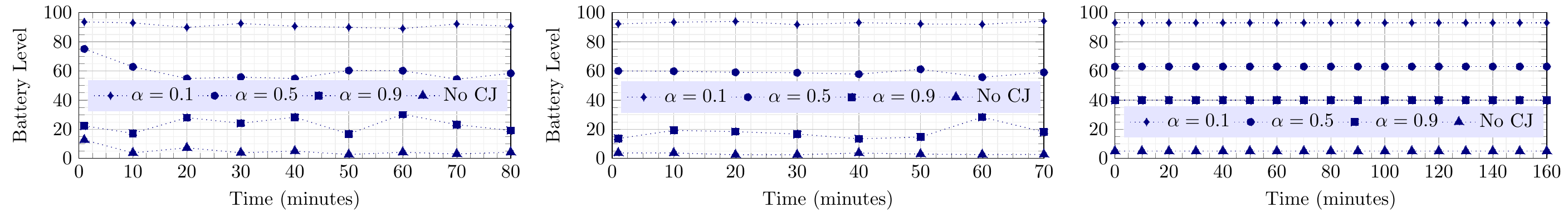}\vspace{-5mm}
\caption{CPU usage across Windows, Linux, and Android. Notice that the Windows device is impacted the most. }
\vspace{-4mm}
\label{expriment:cpu}
\end{centering}
\end{figure*}

\BfPara{\em Results} We obtained two sets of results for each website, with and without cryptojacking. In~\autoref{fig:dynamic}, we plot four test samples obtained from our experiment to demonstrate the behavior of websites with and without cryptojacking. From those results, we observe that loading a website initially consumes significant CPU power (shaded region) in both cases. Once the website is loaded, the CPU consumption decays if the {\em JavaScript} is disabled, indicating no cryptojacking. When {\em JavaScript} is enabled, the CPU consumption is high, indicating cryptojacking. It can also be observed in~\autoref{fig:dynamic}, that the CPU usage varied across the websites, indicating the usage of the throttling parameter highlighted above. The same behavior as with {\em JavaScript} disabled is exhibited when loading a page with {\em JavaScript} that is either benign or of other types of maliciousness than cryptojacking. Through this experiment, we found that cryptojacking consumes anywhere between 10 and 20 times the processing power compared to not using cryptojacking on the same host. To further understand the impact of throttling on CPU usage in different platforms, we conducted another measurement where we used $\alpha=\{0.1, 0.5, 0.9\}$ with the different testing machines. We found a consistent pattern whereby the relationship between $\alpha$ and the CPU usage is linear, as demonstrated in~\autoref{expriment:cpu}.

\begin{figure*}[t]
\centering
\includegraphics[ width=0.9\textwidth]{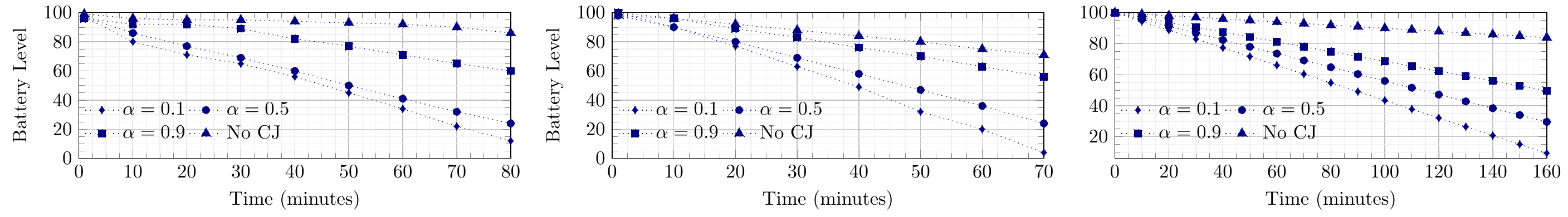}\vspace{-5mm}
\caption{Battery usage across Windows, Linux, and Android, respectively.  Notice that Windows OS is mostly affected by cryptojacking. }
\label{expriment:battery}
\vspace{-4mm}
\end{figure*}

\BfPara{Battery Usage} High CPU usage translates to higher power consumption and quicker battery drainage. To investigate how cryptojacking affects battery drainage, we conducted several experiments using various $\alpha$ values for the various platforms. Here we are interested in the order of battery drainage from a baseline rather than comparing various platforms. The batteries of the different machines are as follows: 65 watt-hours for Windows, 41 watt-hours for Linux, and $\approx$9.88\% watt-hour for Android. 

\BfPara{Memory Usage} In addition to analyzing CPU and battery usage, we also investigated the effect of cryptojacking on the memory usage of Windows, Linux, and Mac. We report results in \autoref{expriment:memory}. Our results show that cryptojacking has no significant relationship with the use of memory since memory usage was random for all experiments. For Windows, with no cryptojacking, the memory usage was $\approx$3.5GB. For $\alpha=0.1,0.5,$ and $0.9$, the memory usage was $\approx$3.1, 3.9, and 3.9GB, respectively. In contrast, for Mac, the memory usage was $\approx$8.5GB, irrespective of the throttling parameter $\alpha$. The randomness in results shows that memory footprint is not a good indicator for cryptojacking detection.

\BfPara{\em Results} For each $\alpha\in \{0.1, 0.5, 0.9\}$, and using the different devices, we ran the {\em JavaScript} script on a fully charged battery. We logged the battery level every 30 seconds, as the script ran on each device with the given $\alpha$ value, starting from a fully-charged battery. Finally, we measure the baseline by running our script without the cryptojacking code. The results are shown in~\autoref{expriment:battery}. As expected, with $\alpha=0.1$, corresponding to the lowest throttling and highest CPU usage, the battery drained very quickly, to $\approx$10\% of its capacity within 80 minutes, compared to $\approx$85\% within the same time when not using cryptojacking. The same result is demonstrated for both the Linux laptop and Android phone. We also notice that the relationship between $\alpha$ and the battery drainage is linear. In examining the CPU and battery usage by cryptojacking websites, as shown above, we highlight clear and unique patterns that can be used to identify those websites.

\subsection{Network Usage and Profiling} \label{sec:throt}
Dynamic artifacts are essential to analyze cryptojacking scripts, especially when scripts are obfuscated. To this end, we also explore the network-level artifacts to uncover the operations of cryptojacking services. 

\begin{figure}[t]
\begin{lstlisting}[caption={WebSocket frames exchanged},label={lst:auth}, style=json]%[t]

// auth request from client to server 
{"type": "auth",
    "params": {
    "site_key": "32 characters key",
    "type": "anonymous", "user": null, "goal": 0 }}
// authed response from server to client    
{ "type": "authed",
	"params": {
	"token": "", "hashes": 0 }}
// job request sent by the server to client	
{ "type": "job",
	"params": {
	"job_id": "164698158344253",
	"blob": "152 characters blob string",
	"target": "ffffff00" }}
// submit message by client to server
{ "type": "submit",
	"params": {
	"job_id": "164698158344253", "nonce": "cfe539d3",
	"result": "256-bit hash" }}
// hash_accept sent by server to client
{ "type": "hash_accept",
	"params": {
	"hashes": 256 }}

\end{lstlisting}
\end{figure}

We noticed that during cryptojacking website execution, the {\em JavaScript} code establishes a WebSocket connection with a remote server and performs a bidirectional data transfer. The WebSocket communication can be monitored using traffic analyzers such as {\em Wireshark}. However, a major issue when using traffic analyzers is that browsers encrypt the web traffic during WebSocket communication. Although significant information, such as source, destination, payload size, and request timings, can still be gathered, the actual data transferred remain encrypted, preventing further analysis. To perform a deeper analysis of WebSocket traffic, we examined the actual data frames {\em in the browser} to understand the communication protocol and payload content of WebSocket connection for possible analysis of cryptojacking websites, outlined below. 

When a WebSocket request is initiated, the client sends an {\em auth} message to the server along with the user information, including {\em sitekey}, {\em type}, and {\em user}. The length of {\em auth} message is 112 bytes. The {\em sitekey} parameter is used by the server to identify the actual user who owns the key of the {\em JavaScript} and adds a balance of hashes to the user's account. The server then authenticates the request parameters and responds with {\em authed} message. The {\em authed} message length is 50 Bytes and includes a token and the total number of hashes received from the client's machine. In the {\em authed} message,  the total number of hashes is 0 since the client has not sent any hashes yet. Then, the server sends {\em job} message to the client. The {\em job} message has a length of 234 Bytes with a {\em job\_id}, {\em blob}, and {\em target}. The {\em target} is a function of the current difficulty in the cryptocurrency to be mined. The client then computes hashes on the {\em nonce} and sends a {\em submit} message back to the server, with {\em job\_id}, {\em nonce}, and the resulting hash. The {\em submit} message has a payload length of 156 Bytes. In response to the {\em submit} message, the server sends {\em hash\_accept} message with an acknowledgment and the total number of hashes received during the session. The {\em hash\_accept} message is 48 Bytes long. This is to be noted that once a webpage is refreshed, the WebSocket connection is terminated and restarted. On the other hand, if multiple tabs are opened in the same browser, the WebSocket connection remains unaffected. In~\autoref{tab:websocket}, we provide details about the WebSocket connection during a cryptojacking session. In~\autoref{lst:auth}, we provide the data frames exchanged between the browser and server during the WebSocket session. The data frames are structured in JSON format.  

\section{Economics of Cryptojacking} \label{sec:economic}
In this section, we evaluate the economic feasibility of cryptojacking by extrapolating the results in our dynamic analysis. We look at the economic feasibility from the perspective of a cryptojacking website's owner, intentional cryptojacking, malicious cryptojacking, and website visitors. For cryptojacking, the reward of the website owner or adversary depends on the number of hashes produced when a website visitor visits the website.  We formulate the analysis as a feasibility: how much of the energy consumed by cryptojacking scripts (cost) is transferred to the cryptojacking website owner, whether malicious or benign and how that translates as an alternative to online advertisement. 

\begin{table}[t]
\centering
\caption{Messages exchanged between the client and the server during WebSocket connection. Length is measured in bytes.}
\label{tab:websocket}\vspace{-3mm}
\scalebox{0.8}{
\begin{tabular}{|l|l|l|c|l|}
\hline
\textbf{Message} & \textbf{Source} & \textbf{Sink} & \textbf{Length} & \textbf{Parameters}\\ \hline
{auth}      & client            & server        & 112                   & sitekey, type, user     \\ \hline
{authed}    & server          & client          & 50                    & token, hashes \\ \hline
{job}       & server          & client          & 234                   & job\_id, blob, target   \\ \hline
{submit}    & client            & server        & 156                   & job\_id, result  \\ \hline
{hash\_accept} & server          & client          & 48                    & hashes   \\ \hline
\end{tabular}}
\vspace{-3mm}\vspace{-3mm}
\end{table}
\subsection{Analytical Model}\label{sec:ana}
\begin{figure}[t]
\begin{center}
\includegraphics[ width=0.45\textwidth]{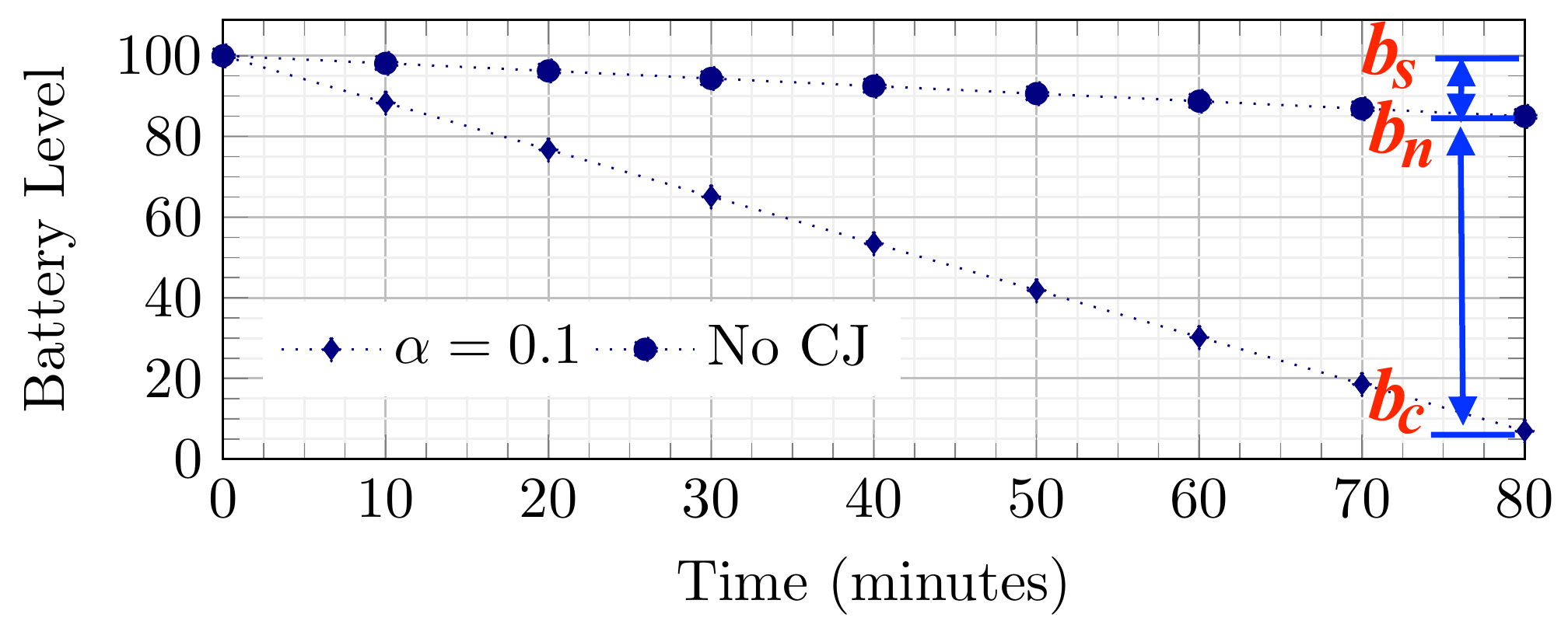}
\vspace{-4mm}
\caption{Battery drain in Windows i7. $b_s$ denotes the starting point of the battery, $b_n$ the normal 80 minutes battery drainage without cryptojacking and $b_c$ denotes the battery drainage with maximum cryptojacking.}
\vspace{-5mm}
\label{fig:overviewbattery}
\end{center}
\end{figure}

To set a stage for our analysis, in~\autoref{fig:overviewbattery}, we present the results from one sample experiment conducted on Windows i7 machine with a cryptojacking website set to minimum throttling ($\alpha$=0.1), indicating a maximum cryptojacking. In this figure, the region between $b_s$ and $b_n$ is a baseline unrelated to cryptojacking--due to the system's normal operation. On the other hand, the region between $b_n$ and $b_c$ is the battery drainage due to cryptojacking. We refer to the energy loss due to such cryptojacking as $L$ for a given user. To formulate the cost (to users) and benefit (to cryptojacking website), let $P$ be the benefit (profit) during a cryptojacking session of $\Delta t$ minutes, and $h$ be the hash rate of the device in hashes/second. At the time of writing this paper, {\em Coinhive} pays $2,894 \times 10^{-8}$ (XMR; currency unit) for 1 million hashes, where 1 XMR equals 200 USD. Therefore, the profit $P$ in XMR in $\Delta{t}=  t_f - t_s$ ($t_f$ and $t_s$ refer to the finish and start time of a session, respectively) can be computed as:    
\begin{eqnarray} 
  \label{equ:profit}  P (\text{XMR}) = ({2,894 \times 10^{-8} \times h \times \Delta{t}}) / {10^{6}}
\end{eqnarray} 
The average hash rate of our test device was 21 hashes/second. For $\Delta{t}=85$ minutes from~\autoref{fig:overviewbattery}, the profit $P$ earned during the session was $3.19 \times 10^{-6}$ XMR or \$ $6.38 \times 10^{-4}$ USD (\$ $1.06 \times 10^{-5}$ USD/second).  This is the upper bound of profit that the device can make in one battery charge.

\begin{table}[t]
\centering
\caption{Results of cryptojacking with different devices. {Here $\alpha$ is the throttling parameter, $h$ (\%), $\Delta{t}$ (mins), $b_n$ (\%), $b_c$ (\%), $W$ (W/h), $P$ (USD; $\times 10^{-4}$), and $L$ (USD, $\times10^{-3}$) are the parameters obtained from~(\ref{equ:profit}) and~(\ref{equ:loss}). The gap is in USD ($\times10^{-3}$) $T$ is the time in years for each device to mine 1 XMR}. }
\label{tab:eco}\vspace{-3mm}
\scalebox{0.75}{
\begin{tabular}{|c|c|c|c|c|c|c|c|c|c|c|}
\hline
\textbf{Device}  & $\Delta{t}$ & $b_n$ & $\alpha$ & $h$&  $b_c$ & $W$ & $P$ & $L$ & $L-P$ & $T$  \\ \hline
\multirow{3}{*}{\bf{Windows}}& \multirow{3}{*}{85}& \multirow{3}{*}{82} & 0.1& 21   & 10  & 65 & 6.4 & 4.5            & 3.8& 50\\ \cline{4-11} 
&           &         & 0.5        & 14                   & 19          & 65          & 3.1    & 3.7                                                                   & 3.4         & 104                                        \\  \cline{4-11}
                                   &           &         & 0.9        & 5                    & 57          & 65          & 0.44    & 1.6                                                                    & 1.5          & 367                                 \\  \cline{1-11}

\multirow{3}{*}{\bf{Linux}}        & \multirow{3}{*}{71}          & \multirow{3}{*}{70}       & 0.1        & 26                   & 3           & 41          & 6.6    & 5.5                             & 4.8        & 40 \\ \cline{4-11}
                                   &           &         & 0.5        & 16                   & 22          & 41          & 4.1    & 4.2                                                                    & 3.8     & 66  \\ \cline{4-11}
                                   &           &         & 0.9        & 5                    & 54          & 41          & 1.3    & 2.6                                                                   & 2.5        & 214 \\ \cline{1-11}

\multirow{3}{*}{\bf{Android}}        & \multirow{3}{*}{163}          & \multirow{3}{*}{76}       & 0.1        &  5                  & 11           & 9.9          & 2.8    & 0.95                             & 0.67      & 220  \\ \cline{4-11}
                                     &           &         & 0.5        & 3                   & 32          & 9.9                                                & 1.7     & 0.72                               & 0.55      & 369      \\ \cline{4-11}
                                     &           &         & 0.9        & 2                    & 49          & 9.9                                               & 1.1    & 0.54                             & 0.43   & 574        \\ \cline{1-11}

\end{tabular}}\vspace{-3mm}
\end{table}

To calculate $L$, corresponding to battery drainage due to cryptojacking ($b_n-b_c$), we first measure the time it takes to recharge 1\% of the battery and denote it by $t_r$. Therefore, the time required to recover $b_n-b_c$ can be calculated as $t_r \times (b_n-b_c)$. Let $W$ be the power consumed by the laptop to run for one hour and $C$ be the cost of electricity in USD/KWH. Therefore, the loss $L$ in USD for the use of the battery can be computed using:  
\begin{eqnarray} 
  \label{equ:loss}  L (USD)  = C \times W \times t_r \times (b_n-b_c) 
\end{eqnarray}
For our test device, we had the following parameters: $W=65$ watt-hour, $C = 6.418 \times 10^{-5}$ USD/(watt-hour), $b_n$ = 82\% (in~\autoref{fig:overviewbattery}), $b_c$ = 10\% and $t_r$ = 0.015 hour. Thus, the estimated loss during cryptojacking session $L$ was $\approx$ \$$4.5 \times 10^{-3}$ USD, which is seven times the value of $P$, highlighting a big gap cryptojacking{}'s operation model. 

\begin{table}
\centering
\caption{Monthly Profit earned by top websites by applying cryptojacking. {GR denotes global rank, CR denotes the country rank, visits are in Billions, average time duration of visits is in mm-ss, P-CJ is the profit earned by cryptojacking, and P-Ads is revenue earned through ads. ``{---}'' denotes the revenue of the companies that we could not find online.}}
\vspace{-3mm}
\label{tab:topsites}
\scalebox{0.8}{
\begin{tabular}{|l|l|l|l|l|l|l|}
\hline
\textbf{Website}  & \textbf{GR} & \textbf{CR} & \textbf{Visits} & \textbf{Time}  & \textbf{P-CJ} &  \textbf{P-Ads} \\ \hline
google.com         & 1                    & 1                     & 47.09                      & 07:23            & 2.41 M  & 7.94 B              \\ \hline
youtube.com           & 2                    & 2                     & 26.22                      & 20:05              & 3.65 M   & 291 M            \\ \hline
baidu.com                  & 3                    & 1                     & 19.08                      & 08:56              & 1.18 M      & 234 M           \\ \hline
wikipedia.org        & 4                    & 6                     & 6.55                       & 03:51              &  0.17 M    & 160 M          \\ \hline
reddit.com            & 5                    & 4                     & 1.69                       & 10:38              & 0.12 M      &  {---}          \\ \hline
facebook.com         & 6                    & 3                     & 29.87                      & 13:28              & 2.80 M      & 3.3 B          \\ \hline
yahoo.com            & 7                    & 7                     & 5.21                       & 06:19              & 0.22 M    & 250 M            \\ \hline
google.co.in               & 8                    & 1                     & 5.33                       & 07:46              & 0.29 M     & 1.1 B           \\ \hline
qq.com                     & 9                    & 2                     & 3.66                       & 04:02              & 0.10 M     &  {---}           \\ \hline
taobao.com                 & 10                   & 3                     & 1.73                       & 06:25              & 0.08 M     &  {---}          \\ \hline
\end{tabular}}
\vspace{-3mm}
\end{table}

Using the same analysis, we examine if users can use cryptojacking as a source of income. With the same device as above, the number of hashes required to make 1 XMR (\$$200$ USD) is  $3.45 \times 10^{10}$ hashes. Given that the same device generates 21 hashes/second, the time required to make 1 XMR is approximately 52 years, while the energy consumed is many orders of magnitude more costly (note that the calculations here are quite theoretical; to mine 1 XMR, it would take $\approx$321,543 battery charging cycles, each of which would cost  0.41 cent (total of $\approx 1318$). In~\autoref{tab:eco}, we report all the results obtained from the experiment for each device used in for our experiments in the dynamic analysis, along with the amount of time required for each device to mine 1 XMR.

\subsection{Cryptojacking and Online Advertisement} \label{sec:mw}

In-browser cryptojacking is being argued as an alternative to online advertisement. To understand the soundness of this argument, we performed an experiment to analyze and compare the monetary value of in-browser cryptojacking as a replacement for online advertisements. 

We select Alexa's top 10 websites~\cite{Alexa-18}. For each website, we obtained the average number of visitors and the time they spent on those websites during March 2018. Using that and our model from section~\ref{sec:ana} to measure the potential profit those websites could have made using cryptojacking. We assume that visitors on these websites have an average hash rate of 20 hashes/second. We report the results in~\autoref{tab:topsites}, highlighting that those websites would make between \$3.65 million USD (for \textit{youtube.com}) and \$0.10 million USD (\textit{qq.com}) per month.

\begin{table}
\centering
\caption{The estimated monthly earnings. Visits are in millions, the average time of each visit is in mm-ss and the profit (P-CJ) is in USD.}
\label{tab:cjprofit}\vspace{-3mm}
\scalebox{0.8}{
\begin{tabular}{|l|l|l|l|l|r|}
\hline
\textbf{Website} & \textbf{GR} & \textbf{CR} & \textbf{Visits} & \textbf{Time}  & \textbf{P-CJ} \\ \hline
firefoxchina.cn                & 1,088       & 132        & 87.24             & 04:32      & 2,746.9            \\ \hline
baytpbportal.fi                & 1,613       & 591        & 12.16             & 05:36      & 472.9              \\ \hline
mejortorrent.com               & 1,800       & 37         & 22.83             & 04:50      & 766.4             \\ \hline
moonbit.co.in                  & 2,761       & 1,289      & 15.68             & 28:37      & 3,116.5              \\ \hline
shareae.com                    & 3,331       & 1,071       & 5.86              & 04:49      & 196.0               \\ \hline
maalaimalar.com                & 4,090       & 112        & 3.38              & 03:26      & 80.6            \\ \hline
icouchtuner.to                 & 6,084       & 518        & 7.96              & 02:98      & 200.8            \\ \hline
paperpk.com                    & 6,794       & 2,050      & 3.01              & 03:23      & 70.7               \\ \hline
scamadviser.com                & 6,847       & 668        & 4.20              & 02:08      & 62.2              \\ \hline
seriesdanko.to                 & 7,253       & 1,452      & 5.44              & 04:59      & 188.2             \\ \hline
\end{tabular}}\vspace{-3mm}\vspace{-3mm}
\end{table}

Statista~\cite{Statista-17} publishes annual online advertisement revenue reports. We collect the revenues generated by each of those top-10 websites for 2017 (most recent report). We use those figures to examine the potential of cryptojacking as an advertisement alternative at scale. For that, we first obtain a monthly revenue figure for each website by dividing the annual revenue by 12. We compare those numbers to the cryptojacking alternative highlighted above. The results are shown in~\autoref{tab:topsites}, where it can be seen that the revenue earned by operating cryptojacking is negligible compared to the revenue earned through online advertisements. For example, if Google is to switch to cryptojacking, it will make \$2.41 million USD per month. In contrast, Google earns $\approx$\$7.94 Billion USD monthly from online advertisement.

To estimate the revenue by cryptojacking websites, we conducted the same experiment on the top-10 websites in our dataset and computed their estimated profit, shown in~\autoref{tab:cjprofit}. We notice that the maximum profit earned by \textit{firefoxchina} is $\approx$\$2,747 USD. Although the ad revenue for these websites is not available online, we still suspect that \$2,747 USD per month is far too low for a website that has 87.24 million monthly views, each with an average duration of 4 minutes and 32 seconds, as compared to the potential revenues for online advertisement. Those findings align with recent reports indicating that an adversary who compromised 5,000 websites and injected his own cryptojacking scripts could only make $\$$24 USD~\cite{Hern_18}.

We conclude that in-browser cryptojacking is not a feasible alternative for the online advertisement since it generates negligible revenue compared to the existing model. Also, as with most PoW-based systems, the economical analysis of cryptojacking as a model highlights a huge $P$ and $L$ negative gap, making it impractical.

\section{Countermeasures} \label{sec:counter}

\subsection{Existing Countermeasures} \label{sec:excm}
At the browser level, existing countermeasures include web extensions such as No Coin, Anti Miner, and No Mining~\cite{Keramidas_18,Tunghobrens-18,Nomining-18}. Each web extension maintains a list of uniform resource locators (URLs) to block while surfing websites. If a user visits a website that is blacklisted by the extension, the user is notified about cryptojacking. However, we show that blacklisting is ineffective since an adaptive attacker can circumvent detection by creating new links not found in the public list of blacklisted URLs.

We set these extensions up on Chrome and evaluated them on our cryptojacking test website. All the extensions detected cryptojacking by reading the WebSocket requests generated by the website to {\em Coinhive}. However, in the next phase, we removed the binding key of our script shown in~\autoref{lst:coinhive}. Without the key, the website establishes the WebSocket connection but does not perform cryptojacking as it cannot verify itself with the server without the key. However, when we tested that on the extensions, all of them wrongly signaled the presence of active cryptojacking. Since extension-based blacklisting does not read the data frames exchanged between WebSockets. Therefore, even the presence of an outdated key or a broken link is falsely labeled as cryptojacking, which highlights a limitation in the detection approach of existing countermeasures.

\begin{figure}[t]
\centering
\includegraphics[ width=0.4\textwidth]{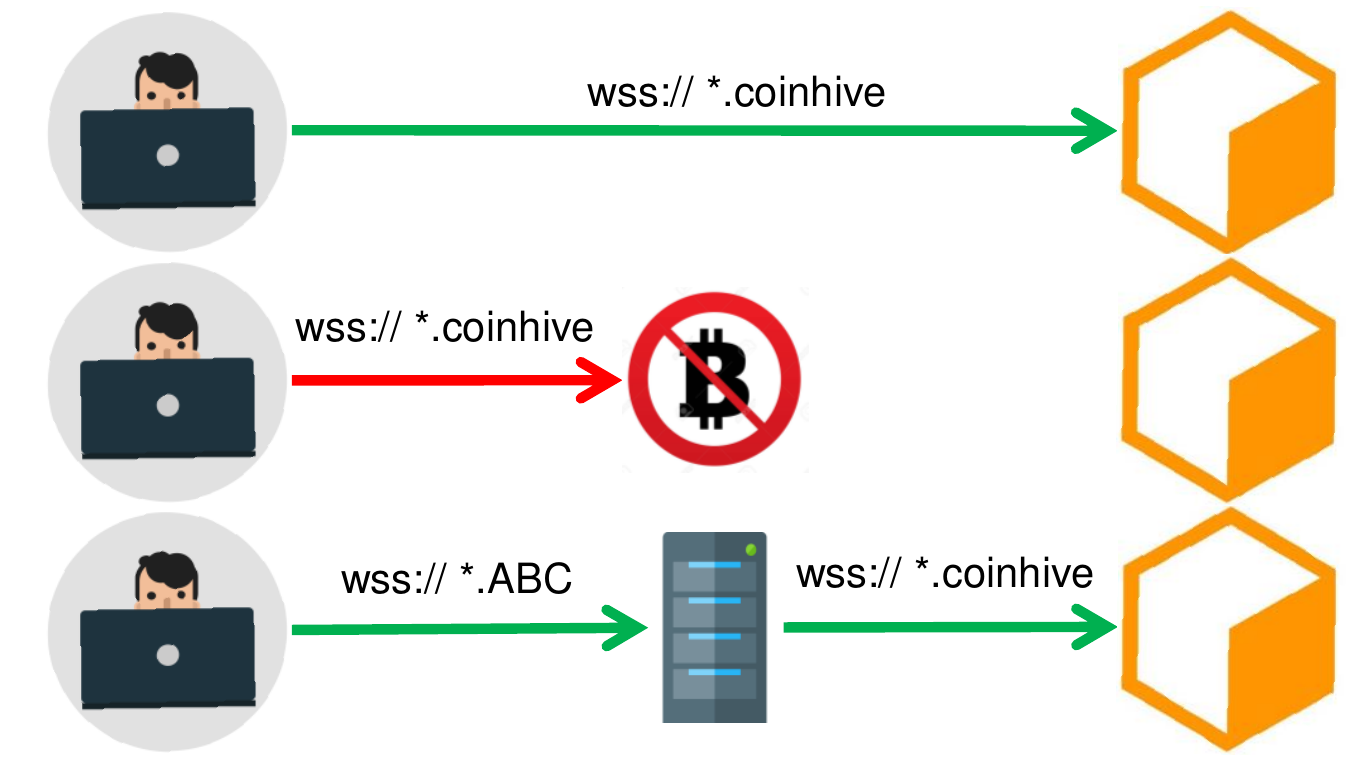}\vspace{-3mm}
\caption{Circumventing cryptojacking detection by relaying WebSocket requests through a third-party proxy server. }
\label{fig:circum}\vspace{-3mm}\vspace{-3mm}
\end{figure}

\BfPara{Evading Detection} \label{sec:evadex}
An attacker, knowing the blacklist, can always evade detection by setting his own third-party server to relay data to and from the cryptojacking server. The cryptojacking website can establish an innocuous WebSocket connection to a third-party server and send data frames and keys to the server. Since anti-cryptojacking extensions will not have the address of a third-party server blacklisted, they will not be able to prevent the connection and cryptojacking. In~\autoref{fig:circum}, we show how an adaptive attacker can circumvent the current countermeasures for cryptojacking. To practically demonstrate that, we set up a test website using {\em Coinhive} script and installed a local relay server. We installed four Chrome extensions blocking the in-browser cryptojacking: No Coin, Anti Miner, No Mining, and Mining Blocker. In the experiment's first phase, we installed the {\em Coinhive} script and ran the website. Each extension detected the WebSocket request and blocked it. To mimic an adaptive attacker, we configured our relay server to act as a proxy, receive socket requests from the browser, and relay them to {\em Coinhive} server. We modified the code in the {\em Coinhive} script and replaced the {\em Coinhive} socket address with our server address. Next, when we visited the website, it started cryptojacking on the client machine. No extension detected it, concluding that it is possible to circumvent the black listing approach for detection through a relay server.

\BfPara{Countering Adaptive Attacker} \label{sec:evadey}
To counter an adaptive attacker and overcome the limitation of existing countermeasures, a better approach is message-based cryptojacking detection in web extensions. Instead of blocking specific URLs, the extensions can monitor the messages exchanged between the user and the server during the cryptojacking sessions. If the messages follow the sequence of web frames illustrated in~\autoref{lst:auth}, the extension can flag them as cryptojacking. This will prevent cryptojacking even if WebSocket requests are relayed through a third party.

To experimentally demonstrate that, we developed a web extension that detects the strings of web frames shown in~\autoref{lst:auth} and notifies the user when the website starts cryptojacking. To test our extension against the existing countermeasures, we deployed a proxy server relaying the data between our test website to the {\em dropzone} server as shown in \autoref{fig:circum}. We installed four Chrome extensions that detect cryptojacking: No Coin, Anti Miner, No Mining, and Mining Blocker. Since all of these extensions take a blacklisting approach for detection, they failed to detect cryptojacking in the presence of the relay server. However, when we installed our newly developed web extension, it immediately flagged cryptojacking upon reading the actual data exchanged between the browser and the relay server. Therefore, we believe the blacklisting approach is insufficient to counter cryptojacking. In contrast, better countermeasures can be developed by deeply inspecting the traffic exchanged between the WebSockets.

\subsection{Adaptive Adversary and Countermeasures} \label{sec:fd}
An adaptive adversary can avoid detection by modifying the cryptojacking script to be similar to benign {\em JavaScript}. Additionally, the adversary can circumvent WebSocket detection through encryption and dummy messages. We note that at the code level, and as shown in our datasets, benign {\em JavaScript} codes are clearly different from cryptojacking codes. Therefore, if an adversary wants to avoid detection, the adversary needs to significantly alter the cryptojacking scripts so as to mix their features with the features of the benign scripts. Given the gap in the feature space between the two classes, as discussed in the paper, an adaptive adversary that tries to mimic the features of another class (i.e., benign features) may be able to do that, but not without sacrificing functional properties of the cryptojacking code which may not be acceptable to the adversary. Cryptojacking scripts are designed to (1) take control of the CPU power, (2) solve PoW challenges, and (3) maintain persistent connections with a {\em dropzone} server to exchange data. These characteristics are quite unique and different from other {\em JavaScript} codes that may simply render an image on the website. Therefore, from a developer's standpoint, writing a cryptojacking script that can perform all such functionalities while giving the same set of code features that are indistinguishable from an image rendering {\em JavaScript} can be difficult to achieve and, therefore, not observed in the wild. 

Similarly, WebSocket-based communication in cryptojacking is different from other WebSocket applications (\ie online chat). One method to bypass detection in this case (also acknowledged in your comments) is by adding an encryption layer or dummy messages that are not detected by the browser extension. Although this is a viable circumvention approach, however, the adversary will (1) bear the encryption cost and (2) circumvent detection in the WebSocket channel only. An intrinsic property of cryptojacking is computing hashes on a {\em nonce} by the victim machine, which can be detected even in the presence of the said circumvention technique. Therefore, resource-based lines of defense can be leveraged to construct more effective countermeasures despite encryption and dummy messages.

\subsection{Discussion} \label{sec:discussion} 

\BfPara{Cryptojacking for Revenue} By showing a negative profit/loss gap, we settle the argument that cryptojacking is not a viable alternative for online advertisement. Moreover, the associated negative reputation may also be a factor in discouraging users from visiting a website that is known to perform cryptojacking on its visitors and we do not see browser-based cryptojacking transforming into a popular and ethical way of generating revenues for online web services. This conclusion is also supported by the low prevalence of cryptojacking sites among the top websites. 

Although the ethical use of cryptojacking is limited, the unethical use may grow as the cryptocurrency market grows and the websites remain vulnerable to {\em JavaScript} injection attacks. Cryptojacking might not be a suitable revenue source for web service providers. However, it may still provide lucrative incentives for adversaries who can make ``easy money'' by compromising vulnerable websites and targeting their visitors. Malicious website owners may combine cryptojacking and online advertisements to increase their overall revenue from website visits.

Our incentive analysis in the paper is first and foremost for understanding the suitability of \cj as employed by some websites as an alternative to existing mechanisms (e.g., advertising). The answer to that question was negative: \cj, when employed for benign purposes as an alternative to ad systems, does not make a lot of sense, for the various gaps it has (cost of mining vs. benefit to the website operator, website operators' benefit from \cj vs traditional ads, and so forth. 

Cryptojacking is not only employed by website operators for revenue but could be injected by adversaries, as has been the case in the overwhelming majority of the websites we studied as well other studies. The findings in our incentive analysis (for benign use) do not apply to the adversary, not even in the slightest sense, for the following reasons. First, any revenue from the cryptojacking for the adversary will be positive, compared to no action (not injecting vulnerable websites with the cryptojacking scripts). Second, cryptojacking attacks are launched solely to abuse visitors' devices on a specific website, thereby influencing the reputation of the website and its ability to attract users and traffic. As a result, we conclude that cryptojacking provides multiple attack avenues for miscreants, and we cannot ignore the potential threat of these attacks or their likelihood of staying prevalent over time. 

\BfPara{Cryptojacking Countermeasures} As shown in \textsection\ref{sec:evadex}, the existing countermeasures for cryptojacking can be easily circumvented. As such, there is a need for strong countermeasures to prevent websites from becoming an attack vector for cryptojacking. Web hosting platforms and ISPs can use the methods outlined in our static analysis (\textsection\ref{sec:static}) to keep a check on the spread of cryptojacking code across websites and notify websites' owners and visitors. 

As a direct result of our dynamic analysis, we argue that web browsers must shield their users from cryptojacking by analyzing the WebSocket payload (\textsection\ref{sec:dynamic}) and reporting fraudulent behavior to the users. We provide a direction towards such improved countermeasures by developing a chrome extension that reads cryptojacking payload during WebSocket communications and notifies the users (\textsection\ref{sec:evadey}).

\section{Related Work} \label{sec:rw}

Concurrent to this work, R{\"u}th \etal \cite{RuthWH18} carried out a measurement study to observe the prevalence of cryptojacking among websites. They obtained blacklisted URLs from the No Coin (\textsection\ref{sec:excm}) web extension and mapped them on a large corpus of websites obtained from the Alexa Top 1M list. In total, they found 1491 suspect websites involved in cryptojacking. However, as shown in \textsection\ref{sec:evadex}, the blacklisting approach to detect and prevent cryptojacking has major limitations and may yield insufficient results to measure prevalence accurately. This perhaps explains the smaller size of their dataset (1491 sites). Concurrently, Eskandari \etal \cite{EskandariLMC18} also examined the prevalence of cryptojacking among websites and used {\em Coinhive} as the most popular platform for cryptojacking. While carried out in parallel to ours, the studies highlight the issue of cryptojacking  through measurements but stop short of conducting any code analysis, detection, and economic analysis.

Huang \etal \cite{HuangDMDGMSWSL14}  were among the first to notice the illegal use of CPU cycles, through malware attacks, for Bitcoin mining. Tahir~\etal~\cite{TahirHDAGZCB17} studied the abuse of virtual machines in cloud services for mining digital currencies. They used micro-architectural execution patterns and CPU signatures to determine if a virtual machine in the cloud was being illegally used for mining purposes and proposed \textit{MineGuard}, a tool to detect mining. Bartino and Nayeem~\cite{BertinoN-17} highlighted worms in IoT devices that hijacked them for mining purposes, pointing to the infamous {\em Linux.Darlloz} worm that hijacked devices running Linux on Intelx86 chip architecture for mining.  Sari and Kilik~\cite{SariS-17}, used Open Source Intelligence (OSINT) to study vulnerabilities in mining pools with the Mirai botnet as a case study. 

Bijamin~\etal~\cite{BijmansBDCCS19} presented a new attack vector where Internet routers were hijacked to launch man-in-the-middle cryptojacking attacks. Another work by Bijamin\etal~\cite{BijmansBD19} analyzed 204 cryptojacking campaigns launched over the Internet and observed that most cryptojacking campaigns were software-based rather than browser-based. Similarly, Pastrana~\etal~\cite{PastranaS19} performed a longitudinal study of the evolution of illicit cryptomining operations over the Internet and uncovered the dynamics of various cryptomining campaigns over the last decade.  Papadopoulos~\etal~\cite{PapadopoulosIM19} examined the impact of in-browser cryptojacking on victim devices and reported that~cryptojacking websites increased the CPU temperature by$\approx$53\% and decreased the CPU performance by up to 57\%. In a similar context, Meland~\etal~\cite{MelandJS19} derived an opposite conclusion to~\cite{PapadopoulosIM19}, stating that a well-configured cryptojacking attack does not harm a user device and may go unnoticed by the users. 

Regarding countermeasures, two notable works have been proposed by Kharraz~\etal~\cite{KharrazMMLMMBAB19} and Hong~\etal~\cite{HongZSLYZMYZH18}. In both studies, the authors applied machine learning techniques to extract code-based features from~cryptojacking websites. Note that some of these works~\cite {PastranaS19,BijmansBD19} were conducted after our initial publication~\cite{SaadKM19}, and the authors have acknowledged our contribution. Others are concurrent works to ours and adapted various approaches to tackle~cryptojacking.  However, our work is unique compared to the existing literature, for it consolidates three major dimensions of in-browser~cryptojacking by performing (uniquely) static, dynamic, and economic analysis. 

\section{Conclusion} \label{sec:conclusion}
We analyze in-browser cryptojacking through the lenses of characterization, static and dynamic analyses, and economic analysis. Our static analysis, applied over 620 websites, unveils unique code complexity characteristics and can be used to detect cryptojacking code from malicious and benign code samples with 100\% accuracy. We explore, through dynamic analysis, how in-browser cryptojacking uses various resources, such as CPU, battery, and network, and use that knowledge to reconstruct the operation of cryptojacking scripts. We also study the economic feasibility of cryptojacking as an alternative to advertising, highlighting its infeasibility. By surveying prior countermeasures and examining their limitations, we highlight long-term solutions, capitalizing on the insights from our static and dynamic analysis and clustering findings. 



\begin{thebibliography}{10}
\providecommand{\url}[1]{#1}
\csname url@samestyle\endcsname
\providecommand{\newblock}{\relax}
\providecommand{\bibinfo}[2]{#2}
\providecommand{\BIBentrySTDinterwordspacing}{\spaceskip=0pt\relax}
\providecommand{\BIBentryALTinterwordstretchfactor}{4}
\providecommand{\BIBentryALTinterwordspacing}{\spaceskip=\fontdimen2\font plus
\BIBentryALTinterwordstretchfactor\fontdimen3\font minus
  \fontdimen4\font\relax}
\providecommand{\BIBforeignlanguage}[2]{{%
\expandafter\ifx\csname l@#1\endcsname\relax
\typeout{** WARNING: IEEEtran.bst: No hyphenation pattern has been}%
\typeout{** loaded for the language `#1'. Using the pattern for}%
\typeout{** the default language instead.}%
\else
\language=\csname l@#1\endcsname
\fi
#2}}
\providecommand{\BIBdecl}{\relax}
\BIBdecl

\bibitem{SaadKM19}
M.~Saad, A.~Khormali, and A.~Mohaisen, ``{Dine and Dash: Static, Dynamic, and
  Economic Analysis of In-browser Cryptojacking},'' in \emph{eCrime}, 2019.

\bibitem{SaadARM21}
\BIBentryALTinterwordspacing
M.~Saad, A.~Anwar, S.~Ravi, and D.~Mohaisen, ``Revisiting nakamoto consensus in
  asynchronous networks: {A} comprehensive analysis of bitcoin safety and
  chainquality,'' in \emph{ACM CCS}, 2021. [Online]. Available:
  \url{https://doi.org/10.1145/3460120.3484561}
\BIBentrySTDinterwordspacing

\bibitem{SaadCM21b}
\BIBentryALTinterwordspacing
M.~Saad, S.~Chen, and D.~Mohaisen, ``Root cause analyses for the deteriorating
  bitcoin network synchronization,'' in \emph{IEEE ICDCS}, 2021. [Online].
  Available: \url{https://doi.org/10.1109/ICDCS51616.2021.00031}
\BIBentrySTDinterwordspacing

\bibitem{SaadCM21}
\BIBentryALTinterwordspacing
------, ``Syncattack: Double-spending in bitcoin without mining power,'' in
  \emph{ACM CCS}, 2021. [Online]. Available:
  \url{https://doi.org/10.1145/3460120.3484568}
\BIBentrySTDinterwordspacing

\bibitem{Scott_18}
\BIBentryALTinterwordspacing
M.~Scott, ``Cryptomining malware fuels most remote code execution attacks,''
  Feb 2018. [Online]. Available: \url{https://tinyurl.com/y9vhrq9w}
\BIBentrySTDinterwordspacing

\bibitem{Zuckerman_18}
\BIBentryALTinterwordspacing
M.~J. Zuckerman, ``Microsoft blocked more than 400,000 malicious cryptojacking
  attempts in one day,'' Apr 2018. [Online]. Available:
  \url{https://tinyurl.com/ya6oj6wm}
\BIBentrySTDinterwordspacing

\bibitem{Slm_18}
\BIBentryALTinterwordspacing
SLM, ``In-browser cryptojacking: What is it and how can you avoid it?'' Jan
  2018. [Online]. Available:
  \url{https://supremelevelmedia.com/browser-cryptojacking-can-avoid/}
\BIBentrySTDinterwordspacing

\bibitem{kerbs}
\BIBentryALTinterwordspacing
B.~Kerbs, ``Who and what is coinhive?'' 2018. [Online]. Available:
  \url{https://krebsonsecurity.com/2018/03/who-and-what-is-coinhive/}
\BIBentrySTDinterwordspacing

\bibitem{coinhive}
\BIBentryALTinterwordspacing
Coinhive, ``Monero {JavaScript} {Mining},'' 2018. [Online]. Available:
  \url{https://coinhive.com/documentation}
\BIBentrySTDinterwordspacing

\bibitem{teamsymantec20}
\BIBentryALTinterwordspacing
TeamSymantec, ``Threat landscape trends – q2 2020.'' [Online]. Available:
  \url{https://symantec-enterprise-blogs.security.com/blogs/threat-intelligence/threat-landscape-trends-q2-2020}
\BIBentrySTDinterwordspacing

\bibitem{Mathur_18}
\BIBentryALTinterwordspacing
N.~Mathur, ``Cybersecurity: Cryptojacking attacks exploded by 8,500\% in 2017,
  says report,'' Apr 2018. [Online]. Available:
  \url{https://tinyurl.com/y84alobt}
\BIBentrySTDinterwordspacing

\bibitem{Singh_18}
\BIBentryALTinterwordspacing
D.~Singh, ``Cryptojacking attacks rose by 8,500\% globally in 2017: report,''
  2018. [Online]. Available: \url{https://tinyurl.com/y9k4ug2q}
\BIBentrySTDinterwordspacing

\bibitem{condliffe_18}
\BIBentryALTinterwordspacing
J.~Condliffe, ``A cryptojacking attack hit thousands of websites,'' 2018.
  [Online]. Available: \url{https://tinyurl.com/ybjck22l}
\BIBentrySTDinterwordspacing

\bibitem{Rayome_18}
\BIBentryALTinterwordspacing
A.~D. Rayome, ``Tesla public cloud environment hacked, attackers accessed
  'non-public' company data,'' 2018. [Online]. Available:
  \url{https://tinyurl.com/y8m79px4}
\BIBentrySTDinterwordspacing

\bibitem{de_18}
\BIBentryALTinterwordspacing
N.~De, ``{UK} cyber security division issues warning on pc 'cryptojacking',''
  Apr 2018. [Online]. Available:
  \url{https://www.coindesk.com/uk-cyber-security-division-issues-warning-on-pc-cryptojacking/}
\BIBentrySTDinterwordspacing

\bibitem{ncsc_18}
\BIBentryALTinterwordspacing
NCSC, ``The cyber threat to uk business 2017-2018 report,'' Apr 2018. [Online].
  Available: \url{https://www.ncsc.gov.uk/cyberthreat}
\BIBentrySTDinterwordspacing

\bibitem{Shaikh_17}
\BIBentryALTinterwordspacing
R.~Shaikh, ``The pirate bay is cryptojacking its visitors' computers to mine
  monero,'' 2017. [Online]. Available: \url{https://tinyurl.com/y9s5mhce}
\BIBentrySTDinterwordspacing

\bibitem{Ernesto_17}
\BIBentryALTinterwordspacing
Ernesto, ``The pirate bay website runs a cryptocurrency miner (updated),'' Sep
  2017. [Online]. Available:
  \url{https://torrentfreak.com/the-pirate-bay-website-runs-a-cryptocurrency-miner-170916/}
\BIBentrySTDinterwordspacing

\bibitem{Jones_2017}
\BIBentryALTinterwordspacing
R.~Jones, ``How to stop pirate bay and other sites from hijacking your cpu to
  mine cryptocoins,'' Sep 2017. [Online]. Available:
  \url{https://tinyurl.com/y9k4ug2q}
\BIBentrySTDinterwordspacing

\bibitem{Zuckerman_2018}
\BIBentryALTinterwordspacing
M.~Zuckerman, ``The ethics of cryptojacking: Rampant malware or ad-free
  internet?'' 2018. [Online]. Available: \url{https://tinyurl.com/yd6u9h39}
\BIBentrySTDinterwordspacing

\bibitem{nakamoto2008bitcoin}
S.~Nakamoto, ``Bitcoin: A peer-to-peer electronic cash system,'' 2008.

\bibitem{SaadSNKSNM20}
\BIBentryALTinterwordspacing
M.~Saad, J.~Spaulding, L.~Njilla, C.~A. Kamhoua, S.~Shetty, D.~Nyang, and
  D.~Mohaisen, ``Exploring the attack surface of blockchain: {A} comprehensive
  survey,'' \emph{{IEEE} Commun. Surv. Tutorials}, vol.~22, no.~3, pp.
  1977--2008, 2020. [Online]. Available:
  \url{https://doi.org/10.1109/COMST.2020.2975999}
\BIBentrySTDinterwordspacing

\bibitem{atozforex}
\BIBentryALTinterwordspacing
A.~Sonewane, ``Top 10 cryptocurrency 2017 | best cryptocurrency to invest,''
  2017. [Online]. Available:
  \url{https://atozforex.com/news/top-10-cryptocurrency-2017/}
\BIBentrySTDinterwordspacing

\bibitem{hileman2017global}
G.~Hileman and M.~Rauchs, ``Global cryptocurrency benchmarking study,''
  \emph{Cambridge Centre for Alternative Finance}, 2017.

\bibitem{bitcoinnews_2017}
\BIBentryALTinterwordspacing
K.~Sedgwick, ``21 statistics that reveal growing demand for the
  cryptocurrency,'' 2017. [Online]. Available: \url{https://goo.gl/BcwAT6}
\BIBentrySTDinterwordspacing

\bibitem{blockexplorer}
\BIBentryALTinterwordspacing
Blockchain, ``Bitcoin block explorer,'' 2018. [Online]. Available:
  \url{https://blockchain.info/}
\BIBentrySTDinterwordspacing

\bibitem{ZimbaWMO20}
\BIBentryALTinterwordspacing
A.~Zimba, Z.~Wang, M.~Mulenga, and N.~H. Odongo, ``Crypto mining attacks in
  information systems: An emerging threat to cyber security,'' \emph{J. Comput.
  Inf. Syst.}, vol.~60, no.~4, pp. 297--308, 2020. [Online]. Available:
  \url{https://doi.org/10.1080/08874417.2018.1477076}
\BIBentrySTDinterwordspacing

\bibitem{Loechner17}
\BIBentryALTinterwordspacing
T.~Loechner, ``Pixalate unveils the list of sites secretly mining
  cryptocurrency,'' 2017. [Online]. Available:
  \url{https://tinyurl.com/y9sbgx92}
\BIBentrySTDinterwordspacing

\bibitem{Netlab360}
\BIBentryALTinterwordspacing
X.~Yang, ``List of top {Alexa} websites with web-mining code embedded on their
  homepage,'' 2017. [Online]. Available: \url{https://tinyurl.com/ybo6u4pf}
\BIBentrySTDinterwordspacing

\bibitem{CalzavaraRB18}
\BIBentryALTinterwordspacing
S.~Calzavara, A.~Rabitti, and M.~Bugliesi, ``Semantics-based analysis of
  content security policy deployment,'' \emph{{TWEB}}, vol.~12, no.~2, pp.
  10:1--10:36, 2018. [Online]. Available: \url{https://doi.org/10.1145/3149408}
\BIBentrySTDinterwordspacing

\bibitem{Similarweb-18}
\BIBentryALTinterwordspacing
SimilarWeb, ``Top websites ranking,'' 2018. [Online]. Available:
  \url{https://www.similarweb.com/top-websites}
\BIBentrySTDinterwordspacing

\bibitem{ZarrasKSHKV14}
A.~Zarras, A.~Kapravelos, G.~Stringhini, T.~Holz, C.~Kruegel, and G.~Vigna,
  ``The dark alleys of madison avenue: Understanding malicious
  advertisements,'' in \emph{{IMC}}, 2014.

\bibitem{Monero}
\BIBentryALTinterwordspacing
M.~Community, ``Monero cryptocurrency,'' 2018. [Online]. Available:
  \url{https://monero.org/}
\BIBentrySTDinterwordspacing

\bibitem{JSEcoin}
\BIBentryALTinterwordspacing
J.~Community, ``{JSECoin}: Digital currency - designed for the web,'' 2018.
  [Online]. Available: \url{https://jsecoin.com/}
\BIBentrySTDinterwordspacing

\bibitem{Wizche17}
Wizsche, ``Malicious javascript dataset,''
  \url{https://github.com/geeksonsecurity/js-malicious-dataset.git}, 2017.

\bibitem{Petrak17}
H.~Petrak, ``Javascript malware collection,''
  \url{https://github.com/HynekPetrak/javascript-malware-collection.git}, 2017.

\bibitem{staff_2017}
\BIBentryALTinterwordspacing
C.~B. Staff, ``21 top examples of javascript,'' 2017. [Online]. Available:
  \url{https://tinyurl.com/y8wqarpb}
\BIBentrySTDinterwordspacing

\bibitem{WatsonMW96}
A.~Watson, T.~J. McCabe, and D.~R. Wallace, \emph{Structured testing: A testing
  methodology using the cyclomatic complexity metric}.\hskip 1em plus 0.5em
  minus 0.4em\relax US Department of Commerce, National Institute of Standards
  and Technology, 1996, vol. 500, no. 235.

\bibitem{FentonN99}
N.~E. Fenton and M.~Neil, ``A critique of software defect prediction models,''
  \emph{IEEE Transactions on software engineering}, vol.~25, no.~5, pp.
  675--689, 1999.

\bibitem{Galeano17}
\BIBentryALTinterwordspacing
S.~R. Galeano, ``On obstructing obscenity obfuscation,'' \emph{{TWEB}},
  vol.~11, no.~2, pp. 12:1--12:24, 2017. [Online]. Available:
  \url{https://doi.org/10.1145/3032963}
\BIBentrySTDinterwordspacing

\bibitem{MohaisenAM15}
\BIBentryALTinterwordspacing
A.~Mohaisen, O.~Alrawi, and M.~Mohaisen, ``{AMAL:} high-fidelity,
  behavior-based automated malware analysis and classification,'' \emph{Comput.
  Secur.}, vol.~52, pp. 251--266, 2015. [Online]. Available:
  \url{https://doi.org/10.1016/j.cose.2015.04.001}
\BIBentrySTDinterwordspacing

\bibitem{KangJMK15}
\BIBentryALTinterwordspacing
H.~Kang, J.~Jang, A.~Mohaisen, and H.~K. Kim, ``Detecting and classifying
  android malware using static analysis along with creator information,''
  \emph{Int. J. Distributed Sens. Networks}, vol.~11, 2015. [Online].
  Available: \url{https://doi.org/10.1155/2015/479174}
\BIBentrySTDinterwordspacing

\bibitem{AlasmaryKAPCAAN19}
H.~Alasmary, A.~Khormali, A.~Anwar, J.~Park, J.~Choi, A.~Abusnaina, A.~Awad,
  D.~Nyang, and A.~Mohaisen, ``Analyzing and detecting emerging internet of
  things malware: {A} graph-based approach,'' \emph{{IEEE} Internet Things J.},
  vol.~6, no.~5, 2019.

\bibitem{MohaisenA13}
\BIBentryALTinterwordspacing
A.~Mohaisen and O.~Alrawi, ``Unveiling zeus: automated classification of
  malware samples,'' in \emph{{WWW}}, 2013, pp. 829--832. [Online]. Available:
  \url{https://doi.org/10.1145/2487788.2488056}
\BIBentrySTDinterwordspacing

\bibitem{Serebrenik11}
A.~Serebrenik, ``Software metrics,''
  \url{http://www.win.tue.nl/~aserebre/2IS55/2010-2011/10.pdf}, 2011.

\bibitem{badge_2016}
\BIBentryALTinterwordspacing
B.~Badge, ``Es-analysis/plato,'' Aug 2016. [Online]. Available:
  \url{https://github.com/es-analysis/plato}
\BIBentrySTDinterwordspacing

\bibitem{LiuCY09}
\BIBentryALTinterwordspacing
J.~Liu, J.~Chen, and J.~Ye, ``Large-scale sparse logistic regression,'' in
  \emph{{ACM} {SIGKDD} International Conference on Knowledge Discovery and Data
  Mining, Paris, France}, 2009, pp. 547--556. [Online]. Available:
  \url{https://doi.org/10.1145/1557019.1557082}
\BIBentrySTDinterwordspacing

\bibitem{Saad20}
\BIBentryALTinterwordspacing
M.~Saad, ``beingmsaad/cryptojacking.'' [Online]. Available:
  \url{https://github.com/beingmsaad/cryptojacking}
\BIBentrySTDinterwordspacing

\bibitem{bruns2009web}
A.~Bruns, A.~Kornstadt, and D.~Wichmann, ``Web application tests with
  selenium,'' \emph{IEEE software}, vol.~26, no.~5, 2009.

\bibitem{seleniumdocumentation}
\BIBentryALTinterwordspacing
S.~Community, ``Selenium browser automation,'' 2018. [Online]. Available:
  \url{https://www.seleniumhq.org/docs/}
\BIBentrySTDinterwordspacing

\bibitem{Alexa-18}
\BIBentryALTinterwordspacing
Alexa, ``The top 500 sites on the websites listed by their 1 month {Alexa}
  traffic rank.'' 2018. [Online]. Available:
  \url{https://www.alexa.com/topsites}
\BIBentrySTDinterwordspacing

\bibitem{Statista-17}
\BIBentryALTinterwordspacing
Statista, ``Google: ad revenue 2001-2017,'' 2018. [Online]. Available:
  \url{https://tinyurl.com/h4rwfyf}
\BIBentrySTDinterwordspacing

\bibitem{Hern_18}
\BIBentryALTinterwordspacing
A.~Hern, ``Huge cryptojacking campaign earns just \$24 for hackers,'' Feb 2018.
  [Online]. Available: \url{https://tinyurl.com/yc5xgvad}
\BIBentrySTDinterwordspacing

\bibitem{Keramidas_18}
\BIBentryALTinterwordspacing
R.~Keramidas, Feb 2018. [Online]. Available:
  \url{https://github.com/keraf/NoCoin}
\BIBentrySTDinterwordspacing

\bibitem{Tunghobrens-18}
\BIBentryALTinterwordspacing
Tunghobrens, ``Anti miner--coin minerblock,'' 2018. [Online]. Available:
  \url{https://tinyurl.com/ybf3jcsj}
\BIBentrySTDinterwordspacing

\bibitem{Nomining-18}
\BIBentryALTinterwordspacing
N.~Mining, ``Secure your browser,'' 2018. [Online]. Available:
  \url{http://www.nomining.com/}
\BIBentrySTDinterwordspacing

\bibitem{RuthWH18}
J.~R{\"{u}}th, T.~Zimmermann, K.~Wolsing, and O.~Hohlfeld, ``Digging into
  browser-based crypto mining,'' in \emph{ACM IMC}, 2018, pp. 70--76.

\bibitem{EskandariLMC18}
\BIBentryALTinterwordspacing
S.~Eskandari, A.~Leoutsarakos, T.~Mursch, and J.~Clark, ``A first look at
  browser-based cryptojacking,'' in \emph{{IEEE} EuroS{\&}P Workshops}, 2018.
  [Online]. Available: \url{https://doi.org/10.1109/EuroSPW.2018.00014}
\BIBentrySTDinterwordspacing

\bibitem{HuangDMDGMSWSL14}
\BIBentryALTinterwordspacing
D.~Y. Huang, H.~Dharmdasani, S.~Meiklejohn, V.~Dave, C.~Grier, D.~McCoy,
  S.~Savage, N.~Weaver, A.~C. Snoeren, and K.~Levchenko, ``Botcoin: Monetizing
  stolen cycles,'' in \emph{ISOC NDSS}, 2014. [Online]. Available:
  \url{https://www.ndss-symposium.org/ndss2014/botcoin-monetizing-stolen-cycles}
\BIBentrySTDinterwordspacing

\bibitem{TahirHDAGZCB17}
R.~Tahir, M.~Huzaifa, A.~Das, M.~Ahmad, C.~A. Gunter, F.~Zaffar, M.~Caesar, and
  N.~Borisov, ``Mining on someone else's dime: Mitigating covert mining
  operations in clouds and enterprises,'' in \emph{RAID}, 2017.

\bibitem{BertinoN-17}
E.~Bertino and N.~Islam, ``Botnets and internet of things security,''
  \emph{Computer}, vol.~50, no.~2, pp. 76--79, 2017.

\bibitem{SariS-17}
A.~Sari and S.~Kilic, ``Exploiting cryptocurrency miners with oisnt
  techniques,'' \emph{Transactions on Networks and Communications}, vol.~5,
  no.~6, 2017.

\bibitem{BijmansBDCCS19}
\BIBentryALTinterwordspacing
H.~L.~J. Bijmans, T.~M. Booij, and C.~Doerr, ``Just the tip of the iceberg:
  Internet-scale exploitation of routers for cryptojacking,'' in \emph{ACM
  CCS}, 2019. [Online]. Available:
  \url{https://doi.org/10.1145/3319535.3354230}
\BIBentrySTDinterwordspacing

\bibitem{BijmansBD19}
\BIBentryALTinterwordspacing
H.~L. Bijmans, T.~M. Booij, and C.~Doerr, ``Inadvertently making cyber
  criminals rich: A comprehensive study of cryptojacking campaigns at internet
  scale,'' in \emph{{USENIX} Security}, 2019. [Online]. Available:
  \url{https://www.usenix.org/conference/usenixsecurity19/presentation/bijmans}
\BIBentrySTDinterwordspacing

\bibitem{PastranaS19}
\BIBentryALTinterwordspacing
S.~Pastrana and G.~Suarez{-}Tangil, ``A first look at the crypto-mining malware
  ecosystem: {A} decade of unrestricted wealth,'' in \emph{ACM IMC}, 2019.
  [Online]. Available: \url{https://doi.org/10.1145/3355369.3355576}
\BIBentrySTDinterwordspacing

\bibitem{PapadopoulosIM19}
\BIBentryALTinterwordspacing
P.~Papadopoulos, P.~Ilia, and E.~P. Markatos, ``Truth in web mining: Measuring
  the profitability and the imposed overheads of cryptojacking,'' in
  \emph{ISC}, 2019. [Online]. Available:
  \url{https://doi.org/10.1007/978-3-030-30215-3\_14}
\BIBentrySTDinterwordspacing

\bibitem{MelandJS19}
\BIBentryALTinterwordspacing
P.~H. Meland, B.~H. Johansen, and G.~Sindre, ``An experimental analysis of
  cryptojacking attacks,'' in \emph{Nordic Conference Secure {IT} Systems},
  2019. [Online]. Available:
  \url{https://doi.org/10.1007/978-3-030-35055-0\_10}
\BIBentrySTDinterwordspacing

\bibitem{KharrazMMLMMBAB19}
\BIBentryALTinterwordspacing
A.~Kharraz, Z.~Ma, P.~Murley, C.~Lever, J.~Mason, A.~Miller, N.~Borisov,
  M.~Antonakakis, and M.~Bailey, ``Outguard: Detecting in-browser covert
  cryptocurrency mining in the wild,'' in \emph{The Web Conference}, 2019.
  [Online]. Available: \url{https://doi.org/10.1145/3308558.3313665}
\BIBentrySTDinterwordspacing

\bibitem{HongZSLYZMYZH18}
G.~Hong, Z.~Yang, S.~Yang, L.~Zhang, Y.~Nan, Z.~Zhang, M.~Yang, Y.~Zhang,
  Z.~Qian, and H.~Duan, ``How you get shot in the back: {A} systematical study
  about cryptojacking in the real world,'' in \emph{ACM CCS}, 2018.

\end{thebibliography}
\end{document}